\documentclass[twocolumn,astrosymb,resetfootnote]{aastex701}

\usepackage{xspace}
\usepackage[export]{adjustbox}
\usepackage{floatrow}

\newcommand{\txs}{TXS\,0506+056\xspace}
\newcommand{\fl}{\textit{Fermi}~LAT\xspace}

\shorttitle{\txs: Hidden ultra-relativistic spine}
\shortauthors{Kovalev et al.}

\graphicspath{{./}{fig/}}
\begin{document}

\title{A hidden ultra-relativistic spine in the jet of a neutrino-associated blazar}

\correspondingauthor{Y.~Y.~Kovalev}
\correspondingauthor{F.~Eppel}
%\email{yykovalev@gmail.com}

\author[0000-0001-9303-3263,gname=Yuri,sname=Kovaev]{Y.~Y.~Kovalev}
\email[show]{yykovalev@gmail.com}
\affiliation{Max-Planck-Institut f\"ur Radioastronomie, Auf dem H\"ugel 69, Bonn D-53121, Germany}

\author[0000-0001-7112-9942,gname=Florian,sname=Eppel]{F.~Eppel}
\email[show]{florian@eppel.space}
\affiliation{Julius-Maximilians-Universität Würzburg, Lehrstuhl f\"ur Astronomie, Emil-Fischer-Straße 31, D-97074 W\"urzburg, Germany}
\affiliation{Max-Planck-Institut f\"ur Radioastronomie, Auf dem H\"ugel 69, Bonn D-53121, Germany}
\affiliation{Joint Institute for VLBI ERIC, Oude Hoogeveensedijk 4, 7991 PD Dwingeloo, The Netherlands}

\author[0000-0002-4431-0890]{D.~C.~Homan}
\email{homand@denison.edu}
\affiliation{Department of Physics and Astronomy, Denison University, Granville, OH 43023, USA}

\author[0000-0003-2914-8554]{A.~V.~Plavin}
\email{alexander@plav.in}
\affiliation{Black Hole Initiative, Harvard University, 20 Garden St, Cambridge, MA 02138, USA}

\author[0009-0006-4186-9978]{P.~Benke}
\email{benke@gfz.de}
\affiliation{GFZ Helmholtz Centre for Geosciences, Telegrafenberg, D-14476 Potsdam, Germany}

\author{A.~K.~Erkenov}
\email{artur-sao@mail.ru}
\affiliation{Special Astrophysical Observatory of the Russian Academy of Sciences, Nizhny Arkhyz 369167, Russia}

\author[0000-0003-4190-7613]{J.~L.~Gómez}
\email{jlgomez@iaa.es}
\affiliation{Instituto de Astrofísica de Andalucía-CSIC, Glorieta de la Astronomía s/n, 18008 Granada, Spain}

\author[0000-0001-5606-6154]{M.~Kadler}
\email{matthias.kadler@astro.uni-wuerzburg.de}
\affiliation{Julius-Maximilians-Universität Würzburg, Lehrstuhl f\"ur Astronomie, Emil-Fischer-Straße 31, D-97074 W\"urzburg, Germany}

\author[0000-0002-0093-4917]{K.~I.~Kellermann}
\email{kkellerm@nrao.edu}
\affiliation{National Radio Astronomy Observatory, 520 Edgemont Road, Charlottesville, VA 22903, USA}

\author[0000-0003-2955-3904]{P.~I.~Kivokurtseva}
\email{kivokurtceva.pi19@physics.msu.ru}
\affiliation{Institute for Nuclear Research, 60th October Anniversary Prospect 7a, Moscow 117312, Russia}

\author[0000-0002-8017-5665]{Yu.~A.~Kovalev}
\email{ykovalev@asc.rssi.ru}
\affiliation{Lebedev Physical Institute, Leninskiy prospekt 53, Moscow 119991, Russia}
\affiliation{Institute for Nuclear Research, 60th October Anniversary Prospect 7a, Moscow 117312, Russia}

\author[0000-0003-1315-3412]{M.~L.~Lister}
\email{mlister@purdue.edu}
\affiliation{Department of Physics and Astronomy, Purdue University, 525 Northwestern Avenue, West Lafayette, IN 47907, USA}

\author[0009-0008-7830-4553]{V.~A.~Makeev}
\email{vmakeev@mpifr-bonn.mpg.de}
\affiliation{Max-Planck-Institut f\"ur Radioastronomie, Auf dem H\"ugel 69, Bonn D-53121, Germany}

\author[0000-0002-0739-700X]{A.~V.~Popkov}
\email{avpopk@gmail.com}
\affiliation{
%Moscow Center for Advanced Studies, Kulakova str. 20, Moscow, 123592, Russia
Moscow Institute of Physics and Technology, Dolgoprudny, Institutsky per., 9, Moscow region, 141700, Russia
}
\affiliation{Lebedev Physical Institute, Leninskiy prospekt 53, Moscow 119991, Russia}
\affiliation{Institute for Nuclear Research, 60th October Anniversary Prospect 7a, Moscow 117312, Russia}

\author[0000-0002-9702-2307]{A.~B.~Pushkarev}
\email{pushkarev.alexander@gmail.com}
\affiliation{Crimean Astrophysical Observatory, 298409 Nauchny, Crimea}
\affiliation{Lebedev Physical Institute, Leninskiy prospekt 53, Moscow 119991, Russia}

\author[0000-0001-9503-4892]{E.~Ros}
\email{ros@mpifr-bonn.mpg.de}
\affiliation{Max-Planck-Institut f\"ur Radioastronomie, Auf dem H\"ugel 69, Bonn D-53121, Germany}

\author[0000-0001-6214-1085]{T.~Savolainen}
\email{tuomas.k.savolainen@aalto.fi}
\affiliation{Aalto University Department of Electronics and Nanoengineering, PL 15500, FI-00076 Aalto, Finland}
\affiliation{Aalto University Metsähovi Radio Observatory, Metsähovintie 114, FI-02540 Kylmälä, Finland}
\affiliation{Max-Planck-Institut f\"ur Radioastronomie, Auf dem H\"ugel 69, Bonn D-53121, Germany}

\author[0000-0001-9172-7237]{Yu.~V.~Sotnikova}
\email{lacerta999@gmail.com}
\affiliation{Special Astrophysical Observatory of the Russian Academy of Sciences, Nizhny Arkhyz 369167, Russia}
\affiliation{Institute for Nuclear Research, 60th October Anniversary Prospect 7a, Moscow 117312, Russia}

\author[0000-0001-6917-6600]{S.~V.~Troitsky}
\email{st@inr.ac.ru}
\affiliation{Institute for Nuclear Research, 60th October Anniversary Prospect 7a, Moscow 117312, Russia}
\affiliation{Physics Department, Lomonosov Moscow State University, 1-2 Leninskie Gory, Moscow 119991, Russia}

\begin{abstract}

Supermassive black holes launch powerful jets of plasma that can accelerate particles to extreme energies, but the physical conditions required to produce high-energy neutrinos remain unknown. In the blazar TXS\,0506+056, the first source individually linked to a high-energy neutrino, radio images had seemed to reveal a jet too slow to sustain the extreme conditions required for neutrino production. Here we resolve this tension using long-term radio monitoring,
particularly Very Long Baseline Interferometry (VLBI) imaging.
We uncover a disturbance in emission propagating with an apparent speed of $21\pm1$ times the speed of light, that is masked by slower, radio-bright features that dominated earlier analyses. We interpret this as the signature of a stratified jet: an ultra-relativistic spine with Lorentz factor $\Gamma>20$ embedded within a slower outer sheath. As the disturbance travels along the spine, it progressively illuminates the sheath, producing the delayed radio flare and naturally accounting for the years-long offset between neutrino and radio emission. The same pattern recurs in a second neutrino-associated event, pointing to a repeatable multi-messenger engine. These findings challenge the standard interpretation of VLBI jet speeds and establish a concrete, testable framework connecting structured jets to the sources of the Universe's highest-energy neutrinos.

\end{abstract}


%\keywords{
%\uat{Blazars}{164} ---
%\uat{Relativistic jets}{1390} ---
%\uat{High Energy astrophysics}{739} ---
%\uat{Neutrino astronomy}{1100}
%}


\section{Introduction} 
\label{s:intro}

The blazar \txs \citep[redshift $z=0.3365$,][]{2018ApJ...854L..32P} rose to prominence in September~2017 when IceCube detected the high‑energy neutrino event IC‑170922A coincident with a bright gamma‑ray flare from the source \citep{0506-scienceGamma}. An archival analysis by IceCube revealed excess neutrino emission between September~2014 and March~2015, even in the absence of a gamma‑ray flare, bolstering the case for \txs as a neutrino source \citep{2018Sci...361..147I}.
This drove extensive multi-wavelength campaigns covering radio through very-high-energy (VHE) bands and their modeling
\citep{0506-scienceGamma,2018ApJ...863L..10A,2018ApJ...864...84K,
0506sheath,2020ApJ...896L..19L,2020ApJ...891..115P,2022ApJ...927..197A,2024MNRAS.527.8784A}.
A second high-energy neutrino from the direction of \txs was subsequently detected by Baikal-GVD in 2021 \citep{2024MNRAS.527.8784A}, further strengthening the case for \txs as a recurring neutrino emitter.

Parsec-scale kinematics measurements at 15~GHz by the MOJAVE (Monitoring of Jets with VLBA Experiments) program suggest a low apparent component pattern speed of $(1.07\pm0.14)c$ \citep{2021ApJ...923...30L}.
Moreover, \citet{2021ApJ...923...67H} estimated its Doppler factor to be only about 2, estimated from the parsec-scale core brightness.
This is in contradiction to the expected high Doppler boosting of VHE gamma-ray and neutrino blazars \citep[e.g.,][]{2007ApJ...664L..71A,2018ApJ...853...68P,2025ApJ...991...33P,2025A&A...700L..12K,2026A&A...706A.351R}.
Multi-messenger simulations for \txs require Doppler factor on the level of 10-20 \citep[e.g.,][]{2018ApJ...864...84K,2019NatAs...3...88G,2022MNRAS.509.2102G}.
A coronal origin for the neutrino emission has been proposed but found to be insufficient to account for the observed flux \citep{2025ApJ...986..104F}, pointing to the jet as the more likely production site.
Subsequent studies revealed a complex parsec-scale morphology and proposed diverse theoretical scenarios for neutrino production \citep[e.g.,][]{2019MNRAS.483L..42K,2019A&A...630A.103B,0506sheath,2020ApJ...896...63L,2025MNRAS.537.3895D}, but none provided a unified account of both the jet kinematics and the observed multi-year VHE-to-radio delay.

We use long‑term MOJAVE 15~GHz monitoring of \txs to study the radio activity associated with the 2017 neutrino--gamma-ray event and to probe the parsec‑scale core and jet during the major flaring event (\autoref{f:mm_lc}). 
While examining the jet model, kinematics and structural evolution, we discovered a very fast relativistic wave propagating along the jet of \txs.
This paper is devoted to the analysis and interpretation of this wave.
Throughout this paper we use the cosmological parameters $\Omega_\mathrm{m} = 0.27$, $\Omega_\Lambda = 0.73$, and $H_0 = 71$~km\,s$^{-1}$~Mpc$^{-1}$ \citep{2009ApJS..180..330K}.
This corresponds to the linear scale of 4.8 pc/mas.

\section{Multi-messenger observations of the flare in \txs}
\label{s:obs}

\begin{figure}
\centering
\includegraphics[width=0.99\linewidth,trim=0cm 0cm 1cm 0cm]{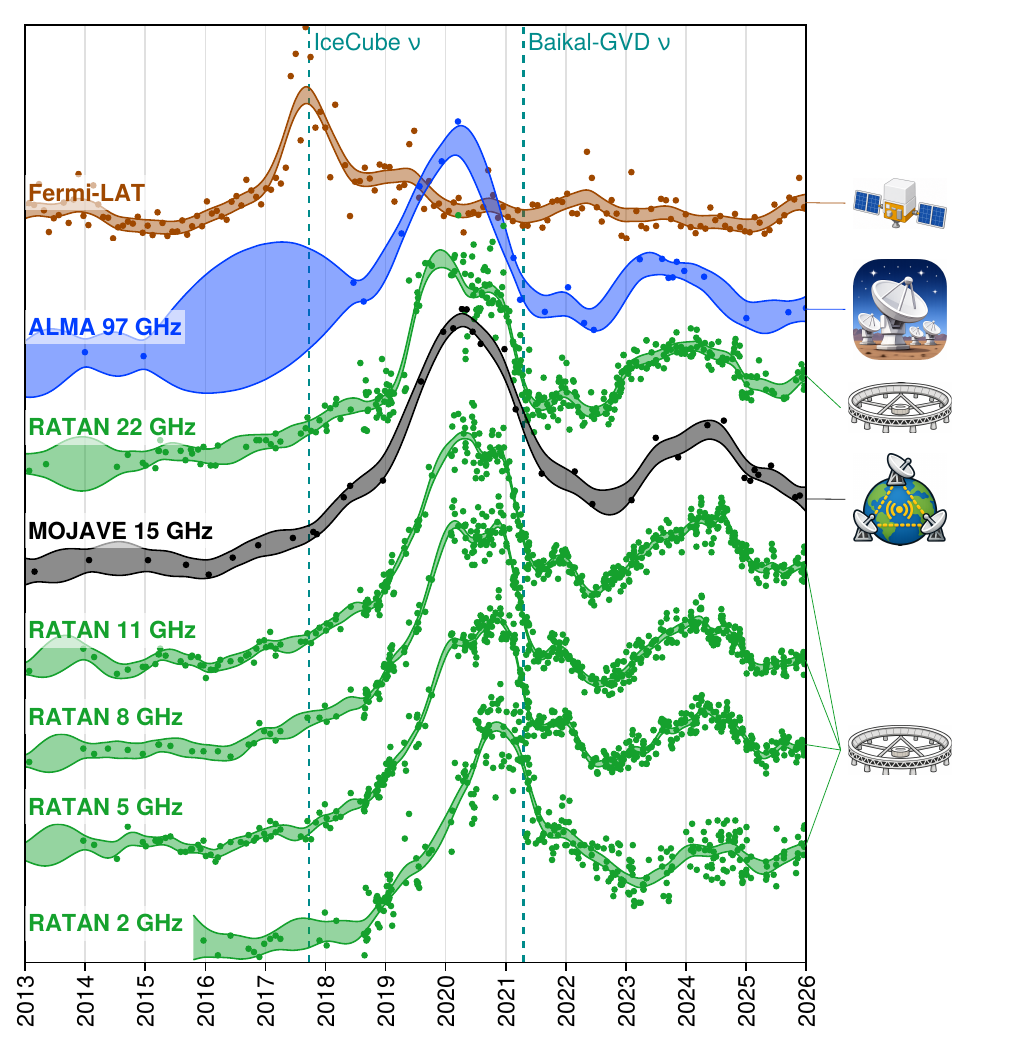}
\caption{Multi-band light curves showing the temporal evolution from gamma-rays to radio frequencies.
Individual measurements are shown as dots.
Smooth shaded bands represent Gaussian-process fits with its $1\sigma$ uncertainty. 
Data uncertainties are discussed in \autoref{s:obs}; they are omitted in the plot to avoid overcrowding, but accounted for in the fits. The vertical axis shows \fl photon flux or radio flux density. The light curves are vertically offset for clarity, so the scaling is arbitrary. Dashed vertical lines mark the arrival times of the IceCube (IC-170922A) and Baikal-GVD (GVD210418CA) neutrino events.
}
\label{f:mm_lc}
\end{figure}

\begin{figure*}[t]
\centering
\includegraphics[width=\textwidth]{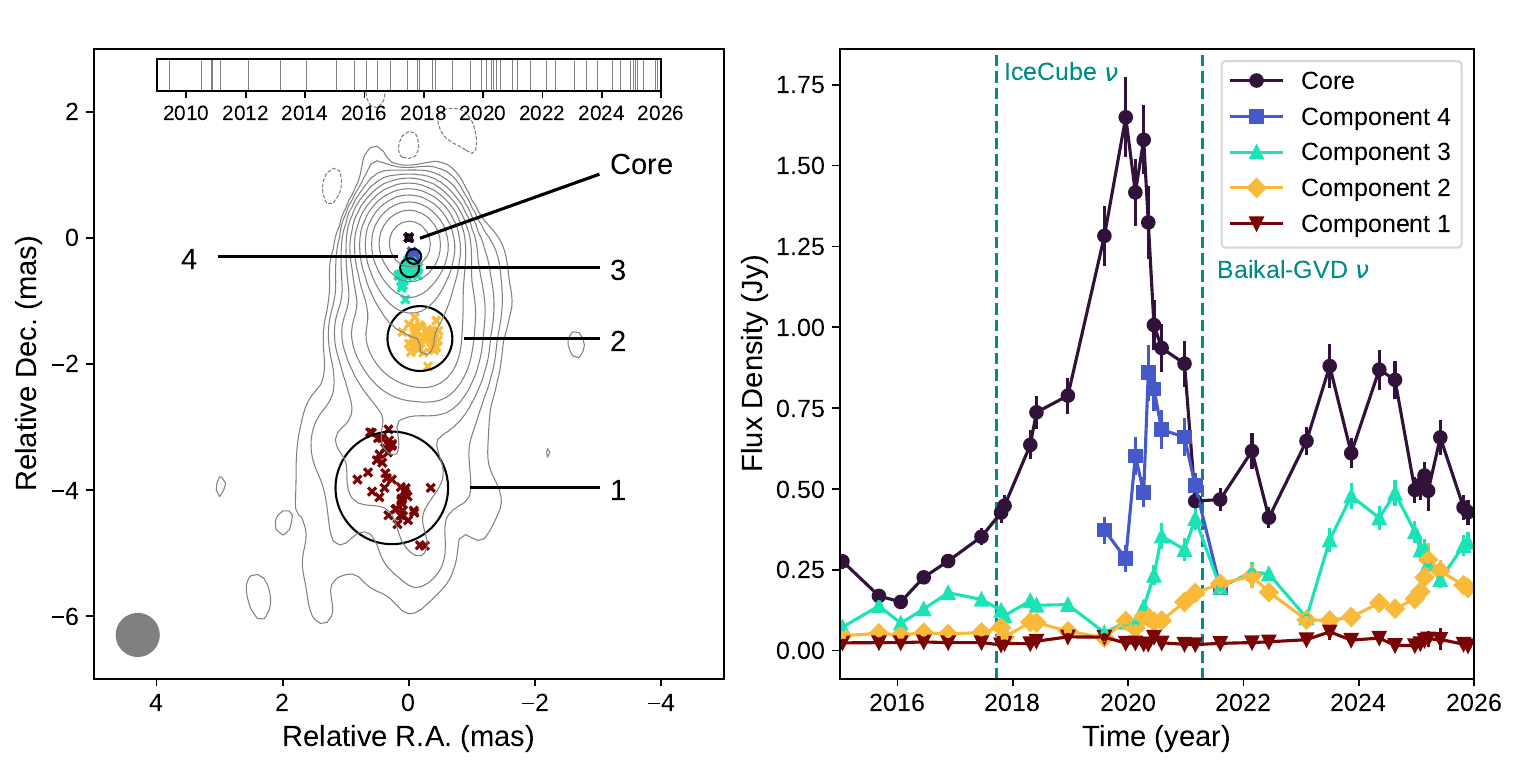}
\caption{
\textit{Left:} Stacked 15~GHz VLBA image of \txs. Component locations for all epochs are shown by x-shaped markers, while the average component positions and sizes are indicated by the black circles. Contours start at 0.41~mJy/beam and increase by factors of two. The restoring beam is shown as a grey circle at the half-power level.
\textit{Right:} Light curves of model-fit components, showing a progressive delay of flare peaks with distance from the core.
\label{f:comp_wave}
}
\end{figure*}

We perform long-term monitoring of AGN with the VLBA at 15~GHz as part of the MOJAVE program\footnote{\url{https://www.cv.nrao.edu/MOJAVE/}} \citep{2018ApJS..234...12L}. \txs was added in 2009 as a gamma-ray bright target and has been observed for 43 epochs through 2025, supplemented by two dedicated epochs in October and November 2017 \citep{0506sheath}. The stacked image compiled from all epochs is shown in \autoref{f:comp_wave} (left panel).

Additionally, we have reduced astrometric/geodetic VLBA observations of \txs at 22~GHz for the time interval 2017-2022 \citep{2023AJ....165..139D} using the \textsc{rPICARD} pipeline \citep{2019A&A...626A..75J}, with hybrid imaging in the \textsc{Difmap} software package \citep{difmap}. 
The limited scan time constrains the dynamic range and amplitude calibration accuracy, but the data quality is sufficient for our purposes.

A major radio flare peaking in 2020 was detected by many facilities \citep[e.g.,][]{OVRO_IC_21,2024MNRAS.527.8784A}, as illustrated in \autoref{f:mm_lc}.
Radio observations of \txs by the transit-mode telescope RATAN-600 \citep{1979S&T....57..324K} are shown for frequencies 2, 5, 8, 11, and 22~GHz with typical uncertainties in total flux density at the 5--10\% level.
Details of the RATAN-600 observations and data processing can be found in \citet{1999A&AS..139..545K, 2018AstBu..73..494T, 2020AdSpR..65..745K}. 
We supplement RATAN single-dish light curves with publicly available ALMA observations at 90--110~GHz from the ALMA Calibrator Source Catalogue\footnote{\url{https://almascience.eso.org/sc/}}. 

\fl photon flux data are taken from the Light Curve Repository \citep{2023ApJS..265...31A} with monthly binning and a fixed photon spectral index of $-2.0$; they cover the energy range 0.1--100~GeV.
We compare the multi-band electromagnetic data with two high-energy neutrino events, associated with \txs:
IceCube event IC‑170922A on 22~September~2017 with an energy estimate of 290~TeV
\citep{2018Sci...361..147I} and Baikal-GVD event GVD210418CA on 18~April~2021 with an energy of 224~TeV \citep{2024MNRAS.527.8784A}.

\begin{figure*}
\centering
\includegraphics[width=1\linewidth,clip,trim=0.5cm 0cm 0.5cm 0cm,valign=c]{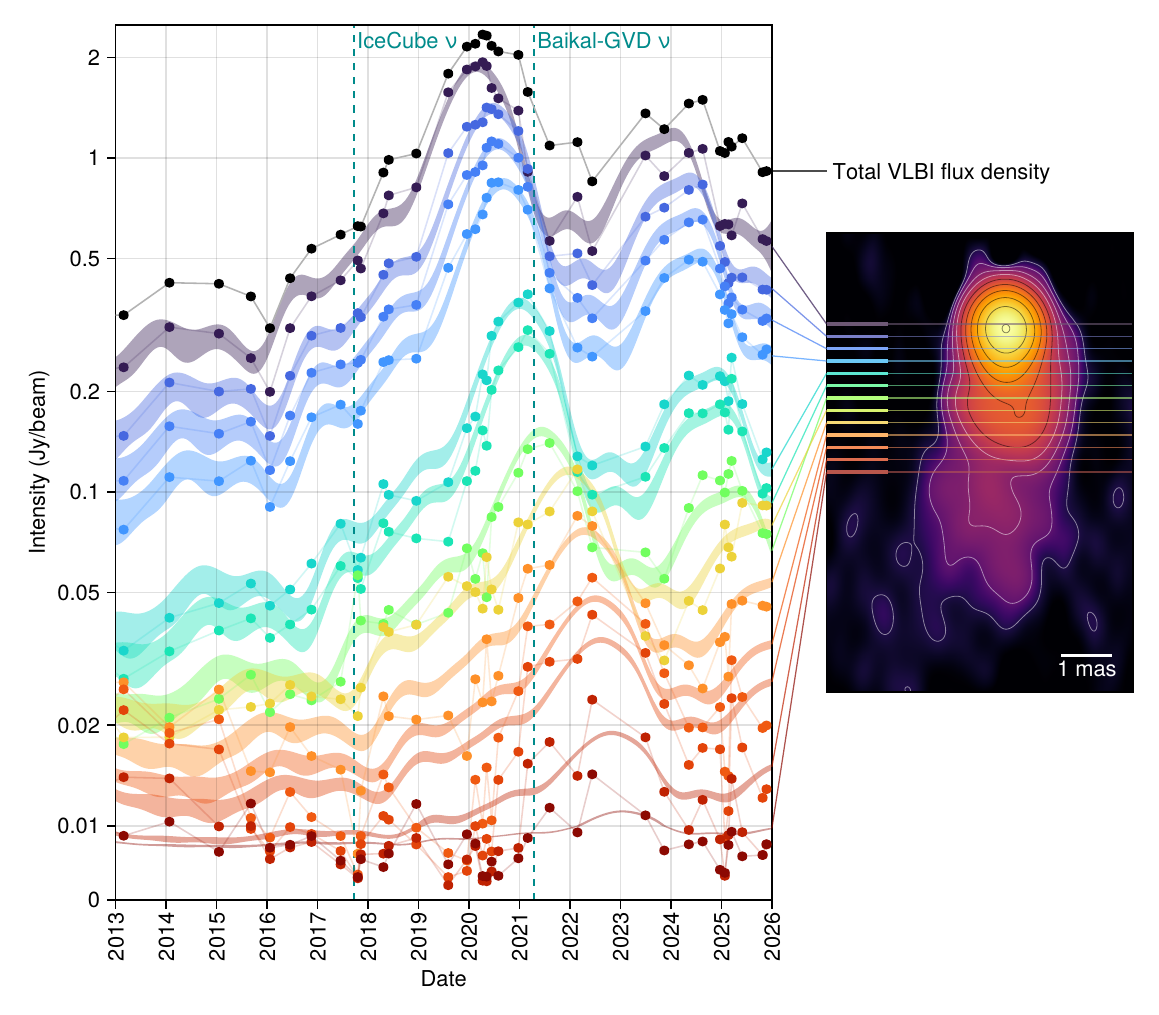}
\caption{
%\footnote{Interactive version of this figure is available at \url{https://aplavin.github.io/txs0506-mojave-wave/time-animation/}.}
\textit{Left:} Stacked MOJAVE VLBA 15~GHz image of \txs with horizontal lines indicating jet slices perpendicular to the jet axis at different core separations. The color of the lines represents distance to the core at both panels.
\textit{Right:} 
Stokes~I intensity light curves at 15~GHz for slices at different core separations, showing the progressive delay at the same radio frequency with distance along the jet. The total VLBI flux density is shown in black (in Jansky, same vertical scale). Dots are individual-epoch measurements; shaded bands show $1\sigma$ uncertainties of a Gaussian-process fit: the exact same function is fit to all light curves, with the time delay and amplitude scale being the only varying parameters. Dashed vertical lines mark the neutrino event times as in \autoref{f:mm_lc}.
\label{f:wave_slices}
}
\end{figure*}

\section{Fast wave propagation in the parsec-scale jet of \txs}
\label{s:wave}

Long-term VLBA monitoring of \txs reveals a coherent wave of enhanced radio emission propagating downstream along the parsec-scale jet at an apparent speed of about $21c$. The slow or stationary components that dominated previous analyses effectively concealed this fast wave. 
Throughout this paper we use the terms 'wave' or 'disturbance' interchangeably to describe the observed propagating emission enhancement; its physical nature is discussed in \autoref{s:stratification}.
The wave appears consistently across multiple independent, methodologically distinct analyses: visibility-domain model fitting using \textsc{Difmap} \citep{difmap,2021ApJ...923...30L} performed independently by different members of the MOJAVE team, image-plane clustering of CLEAN components (Homan et al., in prep.), and wavelet-based analysis \citep[WISE;][]{2015A&A...574A..67M}. Despite their differing assumptions, all methods reveal the same fast propagating feature alongside several standing or slowly moving components. The consistency of these results demonstrates that the wave is physical rather than an artifact of a particular modeling approach.

A representative example set of light curves for the components obtained by the visibility-based model-fitting is shown in \autoref{f:comp_wave}, with a detailed description of model fitting presented in \cite{0506eppel}. We estimate the propagation speed of this emission from the peak times of the individual component flares and their positions at that time. A linear fit to these data suggests a propagation speed of $\mu=(0.71\pm0.09)$\,mas/yr, i.e., $\beta_\mathrm{app} = (14.8 \pm 1.9)c$.
A fit with an additional quadratic term shows no evidence for acceleration, with the corresponding coefficient consistent with zero at the $2\,\sigma$ level.

\begin{figure}
\centering
\includegraphics[width=1.00\linewidth,trim=0.3cm 0.4cm 0cm 0.1cm]{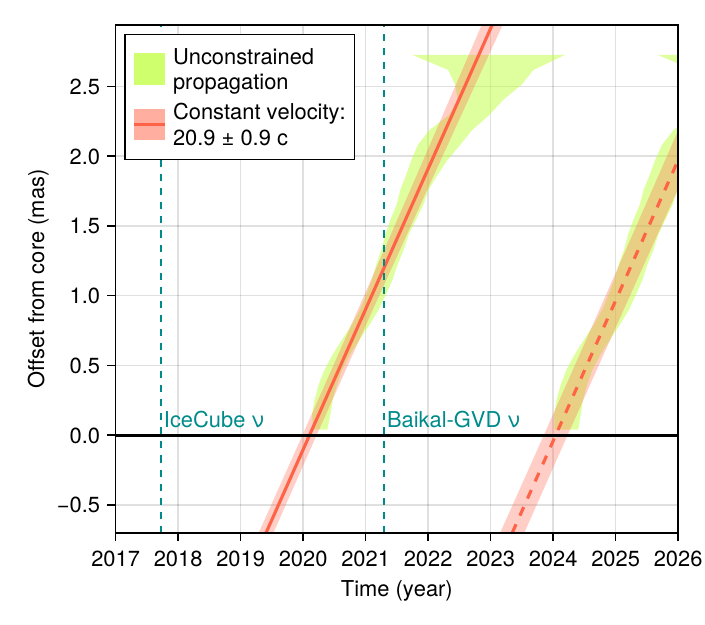}
\caption{
Propagation of two emission waves along the \txs jet at 15~GHz. The green shading shows the flare peak time at each core separation from the unconstrained-delay fit, see \autoref{f:wave_slices} for a subset of slices. The red lines represent a constant-velocity wave propagation,
with the speed determined globally from the joint fit to the light curves at all core separations. The anchor point for the first ($\sim$2020, solid) and second ($\sim$2024, dashed) waves is set at each flare's peak. Vertical dashed lines mark the neutrino detections by IceCube and Baikal-GVD.
}
\label{f:wave_speed}
\end{figure}

To go beyond the small number of model components and avoid model-dependent assumptions, we developed an independent approach based on image slices across the jet. We use MOJAVE VLBA 15~GHz images convolved with a circular beam of 0.7~mas and slice them perpendicular to the average jet axis at a range of distances from the core. For each slice and each observing epoch, we transversely integrate the restored Stokes I image with a beam-area normalization that makes the result independent of the integration extent. This defines a brightness profile along the jet, forming a light curve at each offset from the core. See \autoref{f:wave_slices} for a representative subset of these slices.
These light curves are jointly fitted using Gaussian-process (GP) regression, treating each slice as a time-delayed, amplitude-scaled version of a common underlying flux evolution pattern.
We first fit a model with free unconstrained delays per slice. The resulting delay--separation relation is shown in \autoref{f:wave_speed}.

Both from the visual appearance of the individual light curves (\autoref{f:wave_slices}) and from the delay--separation relation (\autoref{f:wave_speed}), the delays are approximately proportional to the core separation, indicating a constant apparent propagation speed. To estimate the speed most directly, we refit the GP model with the delays $\tau$ constrained to be proportional to the core separation $r$: $\tau = r / \mu$ with $\mu$ being the apparent proper motion. The fitted constant-velocity propagation of the wave is shown as straight line in \autoref{f:wave_speed}. Individual curves derived from this linear-delay model are overlaid in \autoref{f:wave_slices}. For the 15~GHz data, this yields $\mu=(1.01 \pm 0.04)$~mas/y or the apparent linear velocity $\beta_\mathrm{app} = (20.9 \pm 0.9)c$. 
This value is consistent within $3\,\sigma$ with the modelfit approach.

We regard the GP-based estimate as the most robust measurement of the wave propagation speed. It exploits the full accessible range of core separations up to about 3~mas, whereas the modelfit components are predominantly concentrated within $\sim$1.5~mas. The GP approach requires no decomposition of the brightness distribution into discrete components; this decomposition is both model-dependent and subjective, particularly in a complex source. Finally, the speed is determined from a simultaneous global fit to all epochs, all separations, and both waves, rather than from peak times of a small number of individual component light curves from a single wave. The mild discrepancy between the two methods is therefore expected.

We stress the striking consistency: a single temporal profile, merely shifted in time and rescaled in amplitude, captures the flux evolution at every distance from the core. See the close match between the model and the data across all separations in \autoref{f:wave_slices}. This indicates that the disturbance propagates through the jet with remarkably little change in its intrinsic shape.

In addition to the dominant wave~1 peaking in 2020, a later wave~2 is also apparent from \autoref{f:wave_slices}, with its peak in the first half of 2024. Note that our GP model fits a single underlying temporal pattern to all slices, with just a single time delay and amplitude scaling per slice. The close agreement between the fitted profiles and the observed data for both waves indicates that both disturbances propagate at the same apparent speed and undergo a similar evolution across the jet. This strongly suggests a common physical nature for the two waves.

As an independent check, we fitted Gaussian profiles to the radial emission distribution of each wave individually and tracked the peak positions over time, yielding $\mu_1 = 1.14 \pm 0.03$~mas/y and $\mu_2 = 1.05 \pm 0.3$~mas/y.
These are consistent with the GP result, confirming that both waves propagate at the same apparent speed.
The joint GP fit, which determines a single propagation speed from both waves simultaneously across all slices, is therefore the most reliable characterization of the wave kinematics.
Additionally, our independent analysis of the 22~GHz VLBA data has confirmed the presence of the wave with independently estimated apparent speed value $\mu=0.96\pm0.06$~mas/y.
%%% Vlad got 1.05 +- 0.02 mas/yr from his separate analysis of wave1

\section{The multi-messenger flare properties}
\label{s:delay}

\subsection{The radio flare onset and VHE-to-radio delay} 

The delay between the neutrino--gamma-ray event and the radio flare is central to our interpretation, so we first establish when the radio flare began.
The exact onset of the major radio flare is difficult to determine robustly (\autoref{f:mm_lc}).
\citet{2019MNRAS.483L..42K} discussed a rise in the total 15~GHz flux density beginning in 2016. Visual inspection of the data in \autoref{f:mm_lc}, together with an analysis of the associated increase in core brightness temperature from \citet{2021ApJ...923...67H}, places the start of the 15~GHz flare in the first half of 2017.

For \txs, the delay between the 2017 neutrino--gamma-ray event and the 15~GHz radio flare or wave~1 is $\Delta t_\mathrm{\nu\gamma\text{-}r}\approx 2.5$~y (\autoref{f:mm_lc}).
Remarkably, the delay between the independent Baikal-GVD neutrino event GVD210418CA and wave~2 crossing the core is comparable, about 3~y (\autoref{f:wave_slices}, \autoref{f:wave_speed}). 
Two independent neutrino detections, separated by 3.5~y, are each followed by a distinct radio wave, making chance coincidence unlikely and pointing to a recurring physical mechanism connecting high-energy neutrino emission to the jet activity upstream of the parsec-scale core.

The delay between peaks in higher and lower energy emission is a very well known observational effect; it is related to synchrotron opacity in the radio domain \citep[e.g.,][]{BK79,2002PASA...19...83K,2015A&A...575A..55A,2019MNRAS.486..430K}.
In other blazars, the typical radio-to-gamma delay is a few months \citep{2018MNRAS.480.5517L,2022MNRAS.510..469K}, far shorter than the 2.5~y observed here.
The neutrino/gamma-radio delay has been addressed in recent publications within a multi-zone scenario with re-dissipation or synchrotron opacity effects \citep[e.g.,][]{stathopoulos2026,kun2026}.

\subsection{The core shift}

A natural explanation of the large VHE-to-radio delay would be an unusually large core shift: the apparent VLBI core is observed at a non-zero distance from the central engine due to synchrotron opacity, with the offset increasing toward lower frequencies \citep{1998A&A...330...79L}. However, due to the complex nature of the relatively short jet in \txs, measuring the core shift proves difficult.

We aligned multifrequency VLBI images at 2, 8, 15, and 22~GHz by requiring a smooth spectral index distribution in the optically thin jet regions \citep{2019MNRAS.485.1822P}. The 15 and 22~GHz images come from the data described in \autoref{s:obs}, while the 2 and 8~GHz images were taken from the Astrogeo collection \citep{{2025ApJS..276...38P}}. 
The images across frequencies are not strictly simultaneous, with inter-epoch gaps of up to several months; while the core shift may vary somewhat with source activity, large variations on such timescales are not expected.
Our analysis gives core shift values typical for blazars. The offset of the 15~GHz core from the jet apex is estimated to be about 0.1~mas, with a conservative upper limit of 0.4~mas, assuming the stable inverse frequency dependence of the shift and the conical geometry \citep{1998A&A...330...79L}. The flare peaks across the 8--100~GHz frequency range all occur around 2020.2, with mutual offsets of at most several months, as illustrated in \autoref{f:mm_lc}, consistent with this estimate \citep{2019MNRAS.486..430K}.

If we assume that the jet plasma was already strongly accelerated during the VHE multi-messenger flare in 2017, the core shift upper limit of 0.4~mas and the measured apparent angular speed $\mu\approx1$~mas/y strongly contradict the delay of $\Delta t_\mathrm{\nu\gamma\text{-}r}=2.5$~y. 
Tracing the wave back in time at constant speed, the disturbance would have originated $\sim2.5$~mas upstream of the 15~GHz core at the time of the neutrino event, placing it well into the counter-jet, which is physically implausible. Furthermore, strong jet acceleration after the neutrino event is not expected, as jet acceleration occurs during the initial launching and collimation phase. We therefore conclude that the large delay cannot be explained by opacity or propagation effects alone, and requires a different physical interpretation.

\subsection{The core size and radio wave duration}

Based on our model-fitting results, the size of the VLBI core of \txs, modeled as a circular Gaussian component, is constrained to be about 0.1~mas or smaller (FWHM, the full width at half maximum; all sizes and durations in this subsection are quoted at this level).
An unresolved intrinsic disturbance moving at the apparent angular speed $\mu$ would cross the core in less than 1~y~--- a timescale set by the 0.7~mas beam rather than the intrinsic core size.
However, the measured width of the radio wave is about 2~y (\autoref{f:wave_slices}), even after conservatively deconvolving the beam. 
This significantly exceeds the crossing time for an unresolved disturbance and 
is instead comparable to $\Delta t_{\nu\gamma\text{-}r}$, suggesting either a long injection timescale or a transverse origin of the wave duration. We show below that the latter interpretation naturally explains all observed timescales within a unified physical picture.

Notably, the wave duration of about 2~y in the observer frame shows no strong dependence on distance from the core, as demonstrated by the almost-constant flux profile across all separations in \autoref{f:wave_slices}. This is unexpected if the duration were set by the expanding jet geometry, and instead suggests that the wave propagates through the jet with remarkably little change in its intrinsic  structure.

\begin{figure*}[tb]
    \centering   \floatbox[{\capbeside\thisfloatsetup{floatwidth=sidefil,capbesideposition={right,top},capbesidewidth=.28\linewidth}}]{figure}
    {\caption{
    Schematic view of the stratified spine-sheath jet in the plane of the sky. A fast ultra-relativistic spine is surrounded by a slower outer sheath. A disturbance propagating along the spine triggers a transverse expansion of enhanced emission into the sheath, forming a conical brightening pattern moving downstream. The neutrino and gamma-ray flare originates near the jet base, while the radio flare peaks only after the disturbance has crossed the sheath, naturally producing the observed multi-year delay.
    \label{f:scheme}}}
    {
    \includegraphics[width=0.70\textwidth,trim=0cm 0cm 0cm 0cm]{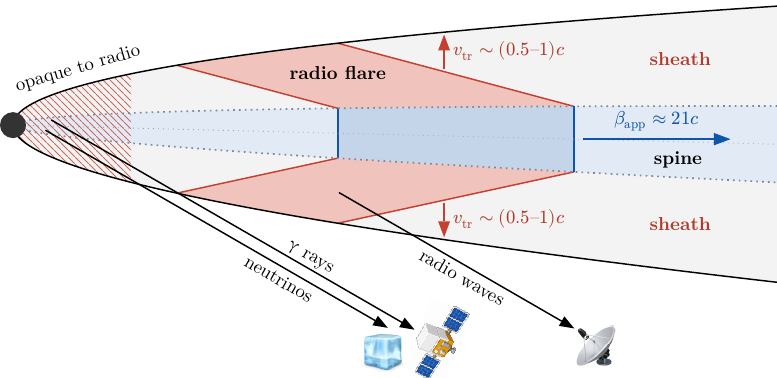}
    }
\end{figure*}

\section{The stratified spine-sheath jet}
\label{s:stratification}

\subsection{Transversely stratified jet scenario}

The observational results reported above find a natural, unified explanation within a transversely stratified jet model. This includes the fast wave propagation, the slow VLBI component motion, the short delay between the VHE multi-messenger flare and the radio flare onset, the large VHE-to-radio delay as well as the long wave duration. Alternatives requiring physically implausible assumptions, such as strong jet acceleration over sub-parsec scales or huge synchrotron opacity, are disfavored as discussed in \autoref{s:delay}.
Spine-sheath jet scenarios of VHE and neutrino production were suggested by, e.g., \citet{2005A&A...432..401G,2014ApJ...793L..18T}.
\citet{2018ApJ...863L..10A} applied this model specifically to explain multi-messenger emission of \txs, while \citet{0506sheath} reported indications of a spine-sheath structure at parsec scales. The companion paper \citet{0506eppel} further reports an EVPA flip across the jet, arising from the interplay between the spine and sheath polarization.

We consider a jet consisting of a fast inner spine surrounded by a slower outer sheath (\autoref{f:scheme}). 
The disturbance originating near the jet base propagates downstream in the spine at a relativistic speed implied by the measured wave velocity of $\beta_\mathrm{app}\approx 21c$. 
In this picture, the neutrino and gamma-ray emission is produced closer to the jet base than the 15~GHz VLBI core, as supported by the synchrotron opacity analysis in \autoref{s:delay}.
The radio emission at tens of GHz is dominated by the sheath rather than the spine. 
This can arise from a deficit of electrons radiating at radio frequencies in the spine, due to either an overall low particle density \citep{1994ApJ...421..153S} or an energy distribution shifted toward higher energies \citep{2005A&A...432..401G}. A weak magnetic field in the spine is another possibility \citep{2005A&A...432..401G, 2008MNRAS.385L..98T}.
%
% YYK: I drop deboosting options since it seem non-feasiable for our parameters (21c and low jet viewing angle).

A shock or fast magnetosonic wave propagating along the spine drives a corresponding disturbance transversely into the sheath, compressing the plasma and amplifying the local magnetic field. The resulting boost in synchrotron emissivity produces the observed transverse expansion of the emission enhancement.
At regions downstream from the core, the radio flux rises as the disturbance progressively energizes the sheath electrons, peaking roughly when the local response is complete.
Because the spine carries the disturbance longitudinally while the sheath merely responds locally, the apparent speed of the radio brightening pattern along the jet reflects the spine propagation speed, not the sheath bulk motion.
The timescale for this transverse propagation is set by the speed at which the disturbance crosses the sheath. For relativistic jet plasmas, sound and Alfvén speeds are generically of the order of
$(0.3-0.6)c$ depending on magnetization and plasma composition, or up to $\sim c$ in the case of a strong relativistic shock \citep[e.g.,][]{2007ApJ...664...26H,2007ApJ...662..835M}.
As a result, the emission enhancement forms a conical pattern moving downstream (\autoref{f:scheme}; see similar patterns in simulations, e.g., \citealt{2016A&A...588A.101F, 2021A&A...647A..77F}).
In this picture, the slowly moving VLBI components \citep[$\lesssim 2c$, see][]{2021ApJ...923...30L,0506eppel} trace the pattern speed of the sheath, not the underlying spine flow; we return to this point in \autoref{s:geometry}.

This setup provides a self-consistent account of the key timescales observed.
The longitudinal propagation of a perturbation in the spine determines the apparent wave velocity measured in \autoref{s:wave} of about $21c$ and the short delay between the neutrino and gamma-ray flare and the onset of the radio flare.
The shorter duration of the gamma-ray flare is also consistent with this picture: the VHE emission originates in the compact region near the jet base, rather than in the extended sheath.
The transverse propagation of the disturbance from the spine through the sheath, at
$\lesssim c$, sets the long VHE-to-radio delay of
$\Delta t_{\nu\gamma\text{-}r} \approx 2.5$~y: the radio flare peaks only once the disturbance has crossed the sheath. The same transverse crossing also determines the observed wave duration of about 2~y (FWHM), since the sheath emission rises and falls as the disturbance passes through it.

The observed flare duration $t_\mathrm{flare,obs}\approx2$~y and a transverse propagation speed of $v_\mathrm{tr}\sim(0.5-1)c$ provide an estimate of the thickness of the sheath in the comoving frame $d_\mathrm{sh} \sim t_\mathrm{flare,obs}\,v_\mathrm{tr}\,\delta_\mathrm{sh}/(1+z)$, where $\delta_\mathrm{sh}$ is the sheath Doppler factor. 
Since the transverse propagation occurs perpendicular to the line of sight, $v_\mathrm{tr}$ is the same in the observer and sheath comoving frames.
For $\delta_\mathrm{sh} \sim 2$ \citep{2021ApJ...923...67H}, this gives $d_\mathrm{sh} \sim (0.5-1)$~pc, broadly consistent with the expected jet geometry. For an intrinsic opening angle assumed as $\alpha_\mathrm{int} = 1.2^\circ$ (see \autoref{s:geometry}) and a spine occupying roughly half the jet radius \citep{2005A&A...432..401G,2006MNRAS.368.1561M}, the sheath thickness at a deprojected distance of $\sim$100~pc from the base is of order 0.5~pc.

The near-constant wave duration along the jet is not trivially expected, since jet expansion would widen the sheath and increase the crossing time with distance. Two scenarios can reconcile this with the observations. The change in sheath width within the 3~mas range probed by our data may simply be too small to produce a measurable variation in duration. Alternatively, the duration may be set by the dissipation timescale of the transverse disturbance rather than the full crossing time, in which case the sheath width estimates above should be treated as lower limits.

The spine-sheath picture also provides a framework for interpreting the neutrino detections, cf.~\citet{2014ApJ...793L..18T,neutrino-from-spine-sheath,2018ApJ...863L..10A}. Two independent neutrino events, IC-170922A at $\sim290$~TeV and GVD210418CA at $\sim224$~TeV, have comparable energies and are each followed by a radio wave propagating at the same apparent speed, with a consistent delay of 2.5 to 3~years. This repeating pattern suggests that both events may originate from the same stationary physical region in the jet \citep[e.g.,][]{Daly_Marscher_1988}, where protons are expected to have intense interactions with ambient photons \citep{Polina}.
This phenomenon should be generic to any jet with strong velocity stratification in which the spine is subdominant in radio emission. A systematic search for similar wave patterns in other sources would help constrain the physical origin of the spine's radio faintness, the mechanism of transverse disturbance propagation, and the prevalence of spine--sheath structure in blazar jets.
We note that the second neutrino event is not accompanied by a detected gamma-ray flare, unlike the first. This could reflect a gamma-opaque or gamma-weak neutrino production zone, since in hadronic models gamma-ray and neutrino emission need not be correlated \citep[e.g.,][]{2018ApJ...864...84K,2019NatAs...3...88G}.

\subsection{Parameters of the stratified jet}
\label{s:geometry}

We have obtained the full apparent opening angle of the \txs jet by applying the method of \citet{MOJAVE_XIV} to the current stacked image (\autoref{f:comp_wave}), which combines 43 epochs spanning 17~y of 15~GHz VLBA observations. The resulting median value is $\alpha_\mathrm{app} = 28.3^\circ \pm 0.6^\circ$. 
The jet width measured in pre-flare epochs (2009--2016) is consistent with that of the full stack, indicating that $\alpha_\mathrm{app}$ traces the structural jet boundary rather than the transient disturbance cone. We therefore take the stacked image as representative of the full sheath width in emission.
We adopt $\alpha_\mathrm{int} = 1.2^\circ$ as the full intrinsic opening angle, the population median for gamma-ray blazars \citep{MOJAVE_XIV}, which yields a jet viewing angle of $\theta=2.4^\circ$.
Combined with the apparent pattern speed $\beta_\mathrm{app}=21c$ (\autoref{s:wave}), this yields Doppler and Lorentz factors $\delta_\mathrm{pattern}=23$ and $\Gamma_\mathrm{pattern}=21$. 
If the pattern is advected with the spine plasma, as assumed in our model, these equal the spine bulk parameters $\delta_\mathrm{spine}$ and $\Gamma_\mathrm{spine}$, which we adopt throughout.

These estimates are dominated by the assumed $\alpha_\mathrm{int}$ and are uncertain at the factor of $\sim$2 level. The inferred geometry and Doppler boosting support the conclusions of \citet{2025ApJ...991...33P,2025A&A...700L..12K}, who associated strongly beamed blazars with neutrino emission.
If the true viewing angle is smaller than adopted here, the inferred Lorentz and Doppler factors would be correspondingly higher.

We note that in this picture the radio emission is dominated by the sheath and is therefore subject to a lower Doppler boosting factor, consistent with the estimates of \citet{2021ApJ...923...67H} based on parsec-scale core brightness. However, the model assumptions adopted in that work may be overly simplistic in the context of a spine-sheath jet structure. 
Furthermore, the apparent kinematics measured from VLBI observations of \txs \citep{2021ApJ...923...30L,0506eppel} should not be directly interpreted as representative of the underlying bulk plasma flow.
The observed standing or slowly moving components in \txs most likely correspond to shocks or plasma instabilities \citep[e.g.,][]{Daly_Marscher_1988,1995ApJ...449L..19G,2000ApJ...533..176H}.

Our results provide the first direct observational evidence that jet velocity stratification resolves the Doppler factor crisis \citep{2010ApJ...722..197L,2018ApJ...853...68P} naturally. The striking discrepancy between the wave propagation speed of $21c$ and the fastest individually identified component speed of $2c$ demonstrates directly that component pattern speeds can systematically underrepresent the true bulk flow in stratified jets, challenging a foundational assumption in AGN jet kinematics. This concern is not about the fidelity of VLBI measurements per se, but about their interpretation in sources with strong velocity stratification. At the same time, the wave speed of $21c$ is not exceptional among blazars \citep{2021ApJ...923...30L}, confirming that VLBI is capable of recovering such speeds when the appropriate analysis is applied.

\section{Summary}

We report a coherent wave of enhanced radio emission propagating down the parsec-scale jet of \txs with an apparent speed of about $21c$, robustly recovered from long-term VLBA imaging with independent analysis approaches. This fast propagation is hidden in conventional component-based kinematics by slowly moving features, but it provides direct evidence for a transversely stratified flow: a radio-faint, ultra-relativistic spine that transports disturbances downstream, surrounded by a slower, radio-bright sheath that responds locally. In this picture, the long delay between the high-energy activity and the radio flare peak is not primarily an opacity effect. It arises naturally because the observable radio brightening is governed by the time required for the disturbance to spread across the sheath, which also explains the wave's nearly constant duration along the jet. The second wave with similar speed and delay points to a repeatable physical mechanism linking neutrino production to the delayed radio activity in the parsec-scale jet.

More broadly, these results offer a concrete observational resolution to the long-standing tension between modest apparent VLBI pattern speeds and the strong Doppler boosting expected in the most extreme blazars, by separating the transport speed of the spine from the radiative dominance of the sheath. This implies that apparent VLBI component speeds do not always reflect the true bulk plasma flow in stratified jets, challenging a foundational assumption in AGN jet kinematics. The model is directly testable: future neutrino activity from \txs and similar sources should be followed by a new fast radio wave with a comparable multi-year delay, and upcoming high-resolution imaging capabilities \citep[e.g.,][]{2023Galax..11..107D,2023Galax..11...92J} should begin to isolate the transverse spine–sheath structure and probe models of neutrino production in stratified jets \citep{neutrino-from-spine-sheath}.

%% Please use the acknowledgment and contribution environments. This will 
%% be anonomyized when the "anonymous" style option is used. 
\begin{acknowledgments}

We thank Mischa Breuhaus, Manel Perucho, Stamatios Stathopoulos, Walter Winter, the MuSES and MOJAVE teams for productive discussions and comments on the manuscript.
This research was funded by the European Union (ERC MuSES project No 101142396).
%Views and opinions expressed are however those of the author(s) only and do not necessarily reflect those of the European Union or the European Research Council. Neither the European Union nor the granting authority can be held responsible for them.
%
E.R.\ was supported by the ERC M2FINDERS project No~101018682.
F.E., M.Ka., and E.R.\ acknowledge support from the Deutsche Forschungsgemeinschaft (DFG, grant 447572188, 443220636 [FOR5195: Relativistic Jets in Active Galaxies]).
A.~V.~Plavin is a postdoctoral fellow at the Black Hole Initiative, which is funded by grants from the John Templeton Foundation (grants 60477, 61479, 62286) and the Gordon and Betty Moore Foundation (grant GBMF-8273). 
TS acknowledges support from the Research Council of Finland projects 362572 and 365088.
The work of A.K.E, P.I.K., Y.A.K., A.V.~Popkov, A.B.P., Y.V.S., and S.V.T. is supported in the framework of the State project ``Science'' by the Ministry of Science and Higher Education of the Russian Federation under the contract 075-15-2024-541.
The views and opinions expressed in this work are those of the authors and do not necessarily reflect the views of these Foundations.

This research made use of the MOJAVE database \citep{2018ApJS..234...12L}, the Astrogeo VLBI FITS image database \citep[DOI: 10.25966/kyy8-yp57,][]{2025ApJS..276...38P}, and the ALMA calibrator data ADS/JAO.ALMA\#2011.0.00001.CAL.
ALMA is a partnership of ESO (representing its member states), NSF (USA) and NINS (Japan), together with NRC (Canada), NSTC and ASIAA (Taiwan), and KASI (Republic of Korea), in cooperation with the Republic of Chile. The Joint ALMA Observatory is operated by ESO, AUI/NRAO and NAOJ.
The National Radio Astronomy Observatory is a facility of the National Science Foundation operated under cooperative agreement by Associated Universities, Inc.
This work is partly based on the data obtained with the RATAN-600 radio telescope at the Special Astrophysical Observatory of the Russian Academy of Sciences (SAO RAS).

\end{acknowledgments}

%%This section gives authors the space to recognize author contributions. The text inside this environment is NOT counted towards the total word quanta. At a minimum, manuscripts are expected to include this text:
%\begin{contribution}
%\yykc{This will be updated (if used at all):} All authors contributed equally.
%\end{contribution}

\facilities{NRAO(VLBA), RATAN-600, ALMA, Fermi(LAT).}

\software{
astropy \citep{2022ApJ...935..167A},
Difmap \citep{difmap},
rPICARD \citep{2019A&A...626A..75J}.
}

\bibliographystyle{aasjournalv7}
\bibliography{0506wave}

@ARTICLE{2026A&A...706A.351R,
       author = {{Rodrigues}, X. and {Rieger}, F. and {Bohdan}, A. and {Padovani}, P.},
        title = "{Hillas meets Eddington: The case for blazars as ultra-high-energy neutrino sources}",
      journal = {\aap},
         year = 2026,
        month = feb,
       volume = {706},
          eid = {A351},
        pages = {A351},
          doi = {10.1051/0004-6361/202556986},
archivePrefix = {arXiv},
       eprint = {2508.18345},
 primaryClass = {astro-ph.HE},
       adsurl = {https://ui.adsabs.harvard.edu/abs/2026A&A...706A.351R}
}

@ARTICLE{2022MNRAS.509.2102G,
       author = {{Gasparyan}, S. and {B{\'e}gu{\'e}}, D. and {Sahakyan}, N.},
        title = "{Time-dependent lepto-hadronic modelling of the emission from blazar jets with SOPRANO: the case of TXS 0506 + 056, 3HSP J095507.9 + 355101, and 3C 279}",
      journal = {\mnras},
         year = 2022,
        month = jan,
       volume = {509},
       number = {2},
        pages = {2102-2121},
          doi = {10.1093/mnras/stab2688},
archivePrefix = {arXiv},
       eprint = {2110.01549},
 primaryClass = {astro-ph.HE},
       adsurl = {https://ui.adsabs.harvard.edu/abs/2022MNRAS.509.2102G}
}

@ARTICLE{2019NatAs...3...88G,
       author = {{Gao}, Shan and {Fedynitch}, Anatoli and {Winter}, Walter and {Pohl}, Martin},
        title = "{Modelling the coincident observation of a high-energy neutrino and a bright blazar flare}",
      journal = {Nature Astronomy},
         year = 2019,
        month = jan,
       volume = {3},
        pages = {88-92},
          doi = {10.1038/s41550-018-0610-1},
archivePrefix = {arXiv},
       eprint = {1807.04275},
 primaryClass = {astro-ph.HE},
       adsurl = {https://ui.adsabs.harvard.edu/abs/2019NatAs...3...88G}
}

@ARTICLE{2025MNRAS.537.3895D,
       author = {{de Gouveia Dal Pino}, E.~M. and {Rodr{\'\i}guez-Ram{\'\i}rez}, J.~C. and {del Valle}, M.~V.},
        title = "{Multimessenger emission from magnetic reconnection in blazar jets: the case of TXS 0506+056}",
      journal = {\mnras},
         year = 2025,
        month = mar,
       volume = {537},
       number = {4},
        pages = {3895-3907},
          doi = {10.1093/mnras/staf251},
archivePrefix = {arXiv},
       eprint = {2411.10210},
 primaryClass = {astro-ph.HE},
       adsurl = {https://ui.adsabs.harvard.edu/abs/2025MNRAS.537.3895D}
}

@ARTICLE{2025ApJ...986..104F,
       author = {{Fiorillo}, Damiano F.~G. and {Testagrossa}, Federico and {Petropoulou}, Maria and {Winter}, Walter},
        title = "{Can the Neutrinos from TXS 0506+056 Have a Coronal Origin?}",
      journal = {\apj},
         year = 2025,
        month = jun,
       volume = {986},
       number = {1},
          eid = {104},
        pages = {104},
          doi = {10.3847/1538-4357/add267},
archivePrefix = {arXiv},
       eprint = {2502.01738},
 primaryClass = {astro-ph.HE},
       adsurl = {https://ui.adsabs.harvard.edu/abs/2025ApJ...986..104F}
}

@misc{kun2026,
      title={Upstream neutrino production and delayed jet emission in the blazar GB6 J1542+6129}, 
      author={Emma Kun and Imre Bartos and Breshna Hadi and Anna Göblyös and Julia Becker Tjus and Peter L. Biermann and Anna Franckowiak and Francis Halzen and Santiago del Palacio and Claudio Ricci},
      year={2026},
      eprint={2605.00785},
      archivePrefix={arXiv},
      primaryClass={astro-ph.HE},
      url={https://arxiv.org/abs/2605.00785}, 
}

@misc{stathopoulos2026,
      title={Delayed Radio Flares in Neutrino-associated Blazars: The Case of TXS 0506+056}, 
      author={S. I. Stathopoulos and C. Yuan and G. Vasilopoulos and F. Testagrossa and D. Karavola and M. Petropoulou and W. Winter},
      year={2026},
      eprint={2604.01196},
      archivePrefix={arXiv},
      primaryClass={astro-ph.HE},
      url={https://arxiv.org/abs/2604.01196}, 
}

@ARTICLE{0506eppel,
       author = {{Eppel}, F. and {Arshakian}, T.~G. and {Benke}, P. and {Degli~Agosti}, C. and {Gómez}, J.~L. and {Homan}, D.~C. and {Kadler}, M. and {Kovalev}, Y.~Y. and {Lister}, M.~L. and {Makeev}, V. and {Plavin}, A.~V. and {Pushkarev}, A.~B. and {Ros}, E. and {Savolainen}, T.},
       title = "{Dissecting the neutrino blazar TXS~0506+056: Polarized jet kinematics and spine-sheath morphology}",
      journal = {\aap, submitted},
         year = 2026
}

@ARTICLE{2023Galax..11...92J,
       author = {{Johnson}, Michael D. and {Doeleman}, Sheperd S. and {G{\'o}mez}, Jos{\'e} L. and {Broderick}, Avery E.},
        title = "{From Vision to Instrument: Creating a Next-Generation Event Horizon Telescope for a New Era of Black Hole Science}",
      journal = {Galaxies},
         year = 2023,
        month = aug,
       volume = {11},
       number = {5},
          eid = {92},
        pages = {92},
          doi = {10.3390/galaxies11050092},
       adsurl = {https://ui.adsabs.harvard.edu/abs/2023Galax..11...92J}
}

@ARTICLE{2023Galax..11..107D,
       author = {{Doeleman}, Sheperd S. and {Barrett}, John and {Blackburn}, Lindy and {Bouman}, Katherine L. and {Broderick}, Avery E. and {Chaves}, Ryan and {Fish}, Vincent L. and {Fitzpatrick}, Garret and {Freeman}, Mark and {Fuentes}, Antonio and {G{\'o}mez}, Jos{\'e} L. and {Haworth}, Kari and {Houston}, Janice and {Issaoun}, Sara and {Johnson}, Michael D. and {Kettenis}, Mark and {Loinard}, Laurent and {Nagar}, Neil and {Narayanan}, Gopal and {Oppenheimer}, Aaron and {Palumbo}, Daniel C.~M. and {Patel}, Nimesh and {Pesce}, Dominic W. and {Raymond}, Alexander W. and {Roelofs}, Freek and {Srinivasan}, Ranjani and {Tiede}, Paul and {Weintroub}, Jonathan and {Wielgus}, Maciek},
        title = "{Reference Array and Design Consideration for the Next-Generation Event Horizon Telescope}",
      journal = {Galaxies},
         year = 2023,
        month = oct,
       volume = {11},
       number = {5},
          eid = {107},
        pages = {107},
          doi = {10.3390/galaxies11050107},
archivePrefix = {arXiv},
       eprint = {2306.08787},
 primaryClass = {astro-ph.IM},
       adsurl = {https://ui.adsabs.harvard.edu/abs/2023Galax..11..107D}
}

@ARTICLE{2006MNRAS.368.1561M,
       author = {{McKinney}, Jonathan C.},
        title = "{General relativistic magnetohydrodynamic simulations of the jet formation and large-scale propagation from black hole accretion systems}",
      journal = {\mnras},
         year = 2006,
        month = jun,
       volume = {368},
       number = {4},
        pages = {1561-1582},
          doi = {10.1111/j.1365-2966.2006.10256.x},
archivePrefix = {arXiv},
       eprint = {astro-ph/0603045},
 primaryClass = {astro-ph},
       adsurl = {https://ui.adsabs.harvard.edu/abs/2006MNRAS.368.1561M}
}

@ARTICLE{1994ApJ...421..153S,
       author = {{Sikora}, Marek and {Begelman}, Mitchell C. and {Rees}, Martin J.},
        title = "{Comptonization of Diffuse Ambient Radiation by a Relativistic Jet: The Source of Gamma Rays from Blazars?}",
      journal = {\apj},
         year = 1994,
        month = jan,
       volume = {421},
        pages = {153},
          doi = {10.1086/173633},
       adsurl = {https://ui.adsabs.harvard.edu/abs/1994ApJ...421..153S}
}

@ARTICLE{2008MNRAS.385L..98T,
       author = {{Tavecchio}, Fabrizio and {Ghisellini}, Gabriele},
        title = "{Spine-sheath layer radiative interplay in subparsec-scale jets and the TeV emission from M87}",
      journal = {\mnras},
         year = 2008,
        month = mar,
       volume = {385},
       number = {1},
        pages = {L98-L102},
          doi = {10.1111/j.1745-3933.2008.00441.x},
archivePrefix = {arXiv},
       eprint = {0801.0593},
 primaryClass = {astro-ph},
       adsurl = {https://ui.adsabs.harvard.edu/abs/2008MNRAS.385L..98T}
}

@ARTICLE{2007ApJ...664...26H,
       author = {{Hardee}, Philip E.},
        title = "{Stability Properties of Strongly Magnetized Spine-Sheath Relativistic Jets}",
      journal = {\apj},
         year = 2007,
        month = jul,
       volume = {664},
       number = {1},
        pages = {26-46},
          doi = {10.1086/518409},
archivePrefix = {arXiv},
       eprint = {0704.1621},
 primaryClass = {astro-ph},
       adsurl = {https://ui.adsabs.harvard.edu/abs/2007ApJ...664...26H}
}

@ARTICLE{2007ApJ...662..835M,
       author = {{Mizuno}, Yosuke and {Hardee}, Philip and {Nishikawa}, Ken-Ichi},
        title = "{Three-dimensional Relativistic Magnetohydrodynamic Simulations of Magnetized Spine-Sheath Relativistic Jets}",
      journal = {\apj},
         year = 2007,
        month = jun,
       volume = {662},
       number = {2},
        pages = {835-850},
          doi = {10.1086/518106},
archivePrefix = {arXiv},
       eprint = {astro-ph/0703190},
 primaryClass = {astro-ph},
       adsurl = {https://ui.adsabs.harvard.edu/abs/2007ApJ...662..835M}
}

@ARTICLE{2016A&A...588A.101F,
       author = {{Fromm}, C.~M. and {Perucho}, M. and {Mimica}, P. and {Ros}, E.},
        title = "{Spectral evolution of flaring blazars from numerical simulations}",
      journal = {\aap},
         year = 2016,
        month = apr,
       volume = {588},
          eid = {A101},
        pages = {A101},
          doi = {10.1051/0004-6361/201527139},
archivePrefix = {arXiv},
       eprint = {1601.03181},
 primaryClass = {astro-ph.HE},
       adsurl = {https://ui.adsabs.harvard.edu/abs/2016A&A...588A.101F}
}

@ARTICLE{2021A&A...647A..77F,
       author = {{Fichet de Clairfontaine}, G. and {Meliani}, Z. and {Zech}, A. and {Hervet}, O.},
        title = "{Flux variability from ejecta in structured relativistic jets with large-scale magnetic fields}",
      journal = {\aap},
         year = 2021,
        month = mar,
       volume = {647},
          eid = {A77},
        pages = {A77},
          doi = {10.1051/0004-6361/202039654},
archivePrefix = {arXiv},
       eprint = {2101.06962},
 primaryClass = {astro-ph.HE},
       adsurl = {https://ui.adsabs.harvard.edu/abs/2021A&A...647A..77F}
}

@ARTICLE{2014ApJ...793L..18T,
       author = {{Tavecchio}, Fabrizio and {Ghisellini}, Gabriele and {Guetta}, Dafne},
        title = "{Structured Jets in BL Lac Objects: Efficient PeV Neutrino Factories?}",
      journal = {\apjl},
         year = 2014,
        month = sep,
       volume = {793},
       number = {1},
          eid = {L18},
        pages = {L18},
          doi = {10.1088/2041-8205/793/1/L18},
archivePrefix = {arXiv},
       eprint = {1407.0907},
 primaryClass = {astro-ph.HE},
       adsurl = {https://ui.adsabs.harvard.edu/abs/2014ApJ...793L..18T}
}

@ARTICLE{2005A&A...432..401G,
       author = {{Ghisellini}, G. and {Tavecchio}, F. and {Chiaberge}, M.},
        title = "{Structured jets in TeV BL Lac objects and radiogalaxies.  Implications for the observed properties}",
      journal = {\aap},
         year = 2005,
        month = mar,
       volume = {432},
       number = {2},
        pages = {401-410},
          doi = {10.1051/0004-6361:20041404},
archivePrefix = {arXiv},
       eprint = {astro-ph/0406093},
 primaryClass = {astro-ph},
       adsurl = {https://ui.adsabs.harvard.edu/abs/2005A&A...432..401G}
}

@ARTICLE{1995ApJ...449L..19G,
       author = {{Gomez}, J.~L. and {Marti}, J.~M.~A. and {Marscher}, A.~P. and {Ibanez}, J.~M.~A. and {Marcaide}, J.~M.},
        title = "{Parsec-Scale Synchrotron Emission from Hydrodynamic Relativistic Jets in Active Galactic Nuclei}",
      journal = {\apjl},
         year = 1995,
        month = aug,
       volume = {449},
        pages = {L19},
          doi = {10.1086/309623},
       adsurl = {https://ui.adsabs.harvard.edu/abs/1995ApJ...449L..19G}
}

@ARTICLE{2000ApJ...533..176H,
       author = {{Hardee}, Philip E.},
        title = "{On Three-dimensional Structures in Relativistic Hydrodynamic Jets}",
      journal = {\apj},
         year = 2000,
        month = apr,
       volume = {533},
       number = {1},
        pages = {176-193},
          doi = {10.1086/308656},
       adsurl = {https://ui.adsabs.harvard.edu/abs/2000ApJ...533..176H}
}

@ARTICLE{2023AJ....165..139D,
       author = {{de Witt}, Aletha and {Jacobs}, Christopher S. and {Gordon}, David and {Bietenholz}, Michael and {Nickola}, Marisa and {Bertarini}, Alessandra},
        title = "{The Celestial Reference Frame at K Band: Imaging. I. The First 28 Epochs}",
      journal = {\aj},
         year = 2023,
        month = apr,
       volume = {165},
       number = {4},
          eid = {139},
        pages = {139},
          doi = {10.3847/1538-3881/aca012},
       adsurl = {https://ui.adsabs.harvard.edu/abs/2023AJ....165..139D}
}

@ARTICLE{2019A&A...626A..75J,
       author = {{Janssen}, M. and {Goddi}, C. and {van Bemmel}, I.~M. and {Kettenis}, M. and {Small}, D. and {Liuzzo}, E. and {Rygl}, K. and {Mart{\'\i}-Vidal}, I. and {Blackburn}, L. and {Wielgus}, M. and {Falcke}, H.},
        title = "{rPICARD: A CASA-based calibration pipeline for VLBI data. Calibration and imaging of 7 mm VLBA observations of the AGN jet in M 87}",
      journal = {\aap},
         year = 2019,
        month = jun,
       volume = {626},
          eid = {A75},
        pages = {A75},
          doi = {10.1051/0004-6361/201935181},
archivePrefix = {arXiv},
       eprint = {1905.01905},
 primaryClass = {astro-ph.IM},
       adsurl = {https://ui.adsabs.harvard.edu/abs/2019A&A...626A..75J}
}

@ARTICLE{2015A&A...574A..67M,
       author = {{Mertens}, Florent and {Lobanov}, Andrei},
        title = "{Wavelet-based decomposition and analysis of structural patterns in astronomical images}",
      journal = {\aap},
         year = 2015,
        month = feb,
       volume = {574},
          eid = {A67},
        pages = {A67},
          doi = {10.1051/0004-6361/201424566},
archivePrefix = {arXiv},
       eprint = {1410.3732},
 primaryClass = {astro-ph.IM},
       adsurl = {https://ui.adsabs.harvard.edu/abs/2015A&A...574A..67M}
}

@INPROCEEDINGS{difmap,
       author = {{Shepherd}, M.~C.},
        title = "{Difmap: an Interactive Program for Synthesis Imaging}",
    booktitle = {Astronomical Data Analysis Software and Systems VI},
         year = 1997,
       editor = {{Hunt}, Gareth and {Payne}, Harry},
       series = {Astronomical Society of the Pacific Conference Series},
       volume = {125},
        month = jan,
        pages = {77},
       adsurl = {https://ui.adsabs.harvard.edu/abs/1997ASPC..125...77S}
}

@ARTICLE{2018ApJ...864...84K,
       author = {{Keivani}, A. and {Murase}, K. and {Petropoulou}, M. and {Fox}, D.~B. and {Cenko}, S.~B. and {Chaty}, S. and {Coleiro}, A. and {DeLaunay}, J.~J. and {Dimitrakoudis}, S. and {Evans}, P.~A. and {Kennea}, J.~A. and {Marshall}, F.~E. and {Mastichiadis}, A. and {Osborne}, J.~P. and {Santander}, M. and {Tohuvavohu}, A. and {Turley}, C.~F.},
        title = "{A Multimessenger Picture of the Flaring Blazar TXS 0506+056: Implications for High-energy Neutrino Emission and Cosmic-Ray Acceleration}",
      journal = {\apj},
         year = 2018,
        month = sep,
       volume = {864},
       number = {1},
          eid = {84},
        pages = {84},
          doi = {10.3847/1538-4357/aad59a},
archivePrefix = {arXiv},
       eprint = {1807.04537},
 primaryClass = {astro-ph.HE},
       adsurl = {https://ui.adsabs.harvard.edu/abs/2018ApJ...864...84K}
}

@ARTICLE{2018ApJ...863L..10A,
       author = {{Ansoldi}, S. and {Antonelli}, L.~A. and {Arcaro}, C. and {Baack}, D. and {Babi{\'c}}, A. and {Banerjee}, B. and {Bangale}, P. and {Barres de Almeida}, U. and {Barrio}, J.~A. and {Becerra Gonz{\'a}lez}, J. and {Bednarek}, W. and {Bernardini}, E. and {Berse}, R. Ch. and {Berti}, A. and {Besenrieder}, J. and {Bhattacharyya}, W. and {Bigongiari}, C. and {Biland}, A. and {Blanch}, O. and {Bonnoli}, G. and {Carosi}, R. and {Ceribella}, G. and {Chatterjee}, A. and {Colak}, S.~M. and {Colin}, P. and {Colombo}, E. and {Contreras}, J.~L. and {Cortina}, J. and {Covino}, S. and {Cumani}, P. and {D'Elia}, V. and {Da Vela}, P. and {Dazzi}, F. and {De Angelis}, A. and {De Lotto}, B. and {Delfino}, M. and {Delgado}, J. and {Di Pierro}, F. and {Dom{\'\i}nguez}, A. and {Dominis Prester}, D. and {Dorner}, D. and {Doro}, M. and {Einecke}, S. and {Elsaesser}, D. and {Fallah Ramazani}, V. and {Fattorini}, A. and {Fern{\'a}ndez-Barral}, A. and {Ferrara}, G. and {Fidalgo}, D. and {Foffano}, L. and {Fonseca}, M.~V. and {Font}, L. and {Fruck}, C. and {Gallozzi}, S. and {Garc{\'\i}a L{\'o}pez}, R.~J. and {Garczarczyk}, M. and {Gaug}, M. and {Giammaria}, P. and {Godinovi{\'c}}, N. and {Guberman}, D. and {Hadasch}, D. and {Hahn}, A. and {Hassan}, T. and {Hayashida}, M. and {Herrera}, J. and {Hoang}, J. and {Hrupec}, D. and {Inoue}, S. and {Ishio}, K. and {Iwamura}, Y. and {Konno}, Y. and {Kubo}, H. and {Kushida}, J. and {Lamastra}, A. and {Lelas}, D. and {Leone}, F. and {Lindfors}, E. and {Lombardi}, S. and {Longo}, F. and {L{\'o}pez}, M. and {Maggio}, C. and {Majumdar}, P. and {Makariev}, M. and {Maneva}, G. and {Manganaro}, M. and {Mannheim}, K. and {Maraschi}, L. and {Mariotti}, M. and {Mart{\'\i}nez}, M. and {Masuda}, S. and {Mazin}, D. and {Mielke}, K. and {Minev}, M. and {Miranda}, J.~M. and {Mirzoyan}, R. and {Moralejo}, A. and {Moreno}, V. and {Moretti}, E. and {Neustroev}, V. and {Niedzwiecki}, A. and {Rosillo}, M. Nievas and {Nigro}, C. and {Nilsson}, K. and {Ninci}, D. and {Nishijima}, K. and {Noda}, K. and {Nogu{\'e}s}, L. and {Paiano}, S. and {Palacio}, J. and {Paneque}, D. and {Paoletti}, R. and {Paredes}, J.~M. and {Pedaletti}, G. and {Pe{\~n}il}, P. and {Peresano}, M. and {Persic}, M. and {Pfrang}, K. and {Prada Moroni}, P.~G. and {Prandini}, E. and {Puljak}, I. and {Garcia}, J.~R. and {Rhode}, W. and {Rib{\'o}}, M. and {Rico}, J. and {Righi}, C. and {Rugliancich}, A. and {Saha}, L. and {Saito}, T. and {Satalecka}, K. and {Schweizer}, T. and {Sitarek}, J. and {{\v{S}}nidari{\'c}}, I. and {Sobczynska}, D. and {Stamerra}, A. and {Strzys}, M. and {Suri{\'c}}, T. and {Tavecchio}, F. and {Temnikov}, P. and {Terzi{\'c}}, T. and {Teshima}, M. and {Torres-Alb{\'a}}, N. and {Tsujimoto}, S. and {Vanzo}, G. and {Vazquez Acosta}, M. and {Vovk}, I. and {Ward}, J.~E. and {Will}, M. and {Zari{\'c}}, D. and {Cerruti}, Matteo},
        title = "{The Blazar TXS 0506+056 Associated with a High-energy Neutrino: Insights into Extragalactic Jets and Cosmic-Ray Acceleration}",
      journal = {\apjl},
         year = 2018,
        month = aug,
       volume = {863},
       number = {1},
          eid = {L10},
        pages = {L10},
          doi = {10.3847/2041-8213/aad083},
archivePrefix = {arXiv},
       eprint = {1807.04300},
 primaryClass = {astro-ph.HE},
       adsurl = {https://ui.adsabs.harvard.edu/abs/2018ApJ...863L..10A}
}

@ARTICLE{1998A&A...330...79L,
       author = {{Lobanov}, A.~P.},
        title = "{Ultracompact jets in active galactic nuclei}",
      journal = {\aap},
         year = 1998,
        month = feb,
       volume = {330},
        pages = {79-89},
          doi = {10.48550/arXiv.astro-ph/9712132},
archivePrefix = {arXiv},
       eprint = {astro-ph/9712132},
 primaryClass = {astro-ph},
       adsurl = {https://ui.adsabs.harvard.edu/abs/1998A&A...330...79L}
}

@ARTICLE{2015A&A...575A..55A,
       author = {{Angelakis}, E. and {Fuhrmann}, L. and {Marchili}, N. and {Foschini}, L. and {Myserlis}, I. and {Karamanavis}, V. and {Komossa}, S. and {Blinov}, D. and {Krichbaum}, T.~P. and {Sievers}, A. and {Ungerechts}, H. and {Zensus}, J.~A.},
        title = "{Radio jet emission from GeV-emitting narrow-line Seyfert 1 galaxies}",
      journal = {\aap},
         year = 2015,
        month = mar,
       volume = {575},
          eid = {A55},
        pages = {A55},
          doi = {10.1051/0004-6361/201425081},
archivePrefix = {arXiv},
       eprint = {1501.02158},
 primaryClass = {astro-ph.HE},
       adsurl = {https://ui.adsabs.harvard.edu/abs/2015A&A...575A..55A}
}

@ARTICLE{2023ApJS..265...31A,
       author = {{Abdollahi}, S. and {Ajello}, M. and {Baldini}, L. and {Ballet}, J. and {Bastieri}, D. and {Becerra Gonzalez}, J. and {Bellazzini}, R. and {Berretta}, A. and {Bissaldi}, E. and {Bonino}, R. and {Brill}, A. and {Bruel}, P. and {Burns}, E. and {Buson}, S. and {Cameron}, R.~A. and {Caputo}, R. and {Caraveo}, P.~A. and {Cibrario}, N. and {Ciprini}, S. and {Cristarella Orestano}, P. and {Crnogorcevic}, M. and {Cutini}, S. and {D'Ammando}, F. and {De Gaetano}, S. and {Digel}, S.~W. and {Di Lalla}, N. and {Di Venere}, L. and {Dom{\'\i}nguez}, A. and {Ramazani}, V. Fallah and {Fegan}, S.~J. and {Ferrara}, E.~C. and {Fiori}, A. and {Fleischhack}, H. and {Franckowiak}, A. and {Fukazawa}, Y. and {Fusco}, P. and {Gammaldi}, V. and {Gargano}, F. and {Garrappa}, S. and {Gasbarra}, C. and {Gasparrini}, D. and {Giglietto}, N. and {Giordano}, F. and {Giroletti}, M. and {Green}, D. and {Grenier}, I.~A. and {Guiriec}, S. and {Gustafsson}, M. and {Hays}, E. and {Horan}, D. and {Hou}, X. and {J{\'o}hannesson}, G. and {Kerr}, M. and {Kocevski}, D. and {Kuss}, M. and {Latronico}, L. and {Li}, J. and {Liodakis}, I. and {Longo}, F. and {Loparco}, F. and {Lorusso}, L. and {Lott}, B. and {Lovellette}, M.~N. and {Lubrano}, P. and {Maldera}, S. and {Manfreda}, A. and {Mart{\'\i}-Devesa}, G. and {Mazziotta}, M.~N. and {Mereu}, I. and {Meyer}, M. and {Michelson}, P.~F. and {Mizuno}, T. and {Monzani}, M.~E. and {Morselli}, A. and {Moskalenko}, I.~V. and {Negro}, M. and {Omodei}, N. and {Orlando}, E. and {Ormes}, J.~F. and {Paneque}, D. and {Panzarini}, G. and {Perkins}, J.~S. and {Persic}, M. and {Pesce-Rollins}, M. and {Pillera}, R. and {Porter}, T.~A. and {Principe}, G. and {Racusin}, J.~L. and {Rain{\`o}}, S. and {Rando}, R. and {Rani}, B. and {Razzano}, M. and {Razzaque}, S. and {Reimer}, A. and {Reimer}, O. and {S{\'a}nchez-Conde}, M. and {Parkinson}, P.~M. Saz and {Scargle}, Jeff and {Scotton}, L. and {Serini}, D. and {Sgr{\`o}}, C. and {Siskind}, E.~J. and {Spandre}, G. and {Spinelli}, P. and {Suson}, D.~J. and {Tajima}, H. and {Thompson}, D.~J. and {Torres}, D.~F. and {Valverde}, J. and {Venters}, T. and {Wadiasingh}, Z. and {Wagner}, S. and {Wood}, K.},
        title = "{The Fermi-LAT Lightcurve Repository}",
      journal = {\apjs},
         year = 2023,
        month = apr,
       volume = {265},
       number = {2},
          eid = {31},
        pages = {31},
          doi = {10.3847/1538-4365/acbb6a},
archivePrefix = {arXiv},
       eprint = {2301.01607},
 primaryClass = {astro-ph.HE},
       adsurl = {https://ui.adsabs.harvard.edu/abs/2023ApJS..265...31A}
}

@ARTICLE{2019A&A...630A.103B,
       author = {{Britzen}, S. and {Fendt}, C. and {B{\"o}ttcher}, M. and {Zaja{\v{c}}ek}, M. and {Jaron}, F. and {Pashchenko}, I.~N. and {Araudo}, A. and {Karas}, V. and {Kurtanidze}, O.},
        title = "{A cosmic collider: Was the IceCube neutrino generated in a precessing jet-jet interaction in TXS 0506+056?}",
      journal = {\aap},
         year = 2019,
        month = oct,
       volume = {630},
          eid = {A103},
        pages = {A103},
          doi = {10.1051/0004-6361/201935422},
       adsurl = {https://ui.adsabs.harvard.edu/abs/2019A&A...630A.103B}
}

@ARTICLE{2020ApJ...896...63L,
       author = {{Li}, Xiaofeng and {An}, Tao and {Mohan}, Prashanth and {Giroletti}, Marcello},
        title = "{The Parsec-scale Jet of the Neutrino-emitting Blazar TXS 0506+056}",
      journal = {\apj},
         year = 2020,
        month = jun,
       volume = {896},
       number = {1},
          eid = {63},
        pages = {63},
          doi = {10.3847/1538-4357/ab8f9f},
archivePrefix = {arXiv},
       eprint = {2005.00300},
 primaryClass = {astro-ph.GA},
       adsurl = {https://ui.adsabs.harvard.edu/abs/2020ApJ...896...63L}
}

@ARTICLE{2022ApJ...927..197A,
       author = {{Acciari}, V.~A. and {Aniello}, T. and {Ansoldi}, S. and {Antonelli}, L.~A. and {Arbet Engels}, A. and {Artero}, M. and {Asano}, K. and {Baack}, D. and {Babi{\'c}}, A. and {Baquero}, A. and {Barres de Almeida}, U. and {Barrio}, J.~A. and {Batkovi{\'c}}, I. and {Becerra Gonz{\'a}lez}, J. and {Bednarek}, W. and {Bernardini}, E. and {Bernardos}, M. and {Berti}, A. and {Besenrieder}, J. and {Bhattacharyya}, W. and {Bigongiari}, C. and {Biland}, A. and {Blanch}, O. and {B{\"o}kenkamp}, H. and {Bonnoli}, G. and {Bo{\v{s}}njak}, {\v{Z}}. and {Busetto}, G. and {Carosi}, R. and {Ceribella}, G. and {Cerruti}, M. and {Chai}, Y. and {Chilingarian}, A. and {Cikota}, S. and {Colombo}, E. and {Contreras}, J.~L. and {Cortina}, J. and {Covino}, S. and {D'Amico}, G. and {D'Elia}, V. and {Vela}, P. Da and {Dazzi}, F. and {De Angelis}, A. and {De Lotto}, B. and {Del Popolo}, A. and {Delfino}, M. and {Delgado}, J. and {Mendez}, C. Delgado and {Depaoli}, D. and {Di Pierro}, F. and {Di Venere}, L. and {Do Souto Espi{\~n}eira}, E. and {Dominis Prester}, D. and {Donini}, A. and {Dorner}, D. and {Doro}, M. and {Elsaesser}, D. and {Fallah Ramazani}, V. and {Fari{\~n}a}, L. and {Fattorini}, A. and {Font}, L. and {Fruck}, C. and {Fukami}, S. and {Fukazawa}, Y. and {Garc{\'\i}a L{\'o}pez}, R.~J. and {Garczarczyk}, M. and {Gasparyan}, S. and {Gaug}, M. and {Giglietto}, N. and {Giordano}, F. and {Gliwny}, P. and {Godinovi{\'c}}, N. and {Green}, J.~G. and {Green}, D. and {Hadasch}, D. and {Hahn}, A. and {Hassan}, T. and {Heckmann}, L. and {Herrera}, J. and {Hoang}, J. and {Hrupec}, D. and {H{\"u}tten}, M. and {Inada}, T. and {Iotov}, R. and {Ishio}, K. and {Iwamura}, Y. and {Jim{\'e}nez Mart{\'\i}nez}, I. and {Jormanainen}, J. and {Jouvin}, L. and {Kerszberg}, D. and {Kobayashi}, Y. and {Kubo}, H. and {Kushida}, J. and {Lamastra}, A. and {Lelas}, D. and {Leone}, F. and {Lindfors}, E. and {Linhoff}, L. and {Lombardi}, S. and {Longo}, F. and {L{\'o}pez-Coto}, R. and {L{\'o}pez-Moya}, M. and {L{\'o}pez-Oramas}, A. and {Loporchio}, S. and {Machado de Oliveira Fraga}, B. and {Maggio}, C. and {Majumdar}, P. and {Makariev}, M. and {Mallamaci}, M. and {Maneva}, G. and {Manganaro}, M. and {Mannheim}, K. and {Mariotti}, M. and {Mart{\'\i}nez}, M. and {Mas Aguilar}, A. and {Mazin}, D. and {Menchiari}, S. and {Mender}, S. and {Mi{\'c}anovi{\'c}}, S. and {Miceli}, D. and {Miener}, T. and {Miranda}, J.~M. and {Mirzoyan}, R. and {Molina}, E. and {Moralejo}, A. and {Morcuende}, D. and {Moreno}, V. and {Moretti}, E. and {Nakamori}, T. and {Nava}, L. and {Neustroev}, V. and {Nievas Rosillo}, M. and {Nigro}, C. and {Nilsson}, K. and {Nishijima}, K. and {Noda}, K. and {Nozaki}, S. and {Ohtani}, Y. and {Oka}, T. and {Otero-Santos}, J. and {Paiano}, S. and {Palatiello}, M. and {Paneque}, D. and {Paoletti}, R. and {Paredes}, J.~M. and {Pavleti{\'c}}, L. and {Pe{\~n}il}, P. and {Persic}, M. and {Pihet}, M. and {Prada Moroni}, P.~G. and {Prandini}, E. and {Priyadarshi}, C. and {Puljak}, I. and {Rhode}, W. and {Rib{\'o}}, M. and {Rico}, J. and {Righi}, C. and {Rugliancich}, A. and {Sahakyan}, N. and {Saito}, T. and {Sakurai}, S. and {Satalecka}, K. and {Saturni}, F.~G. and {Schleicher}, B. and {Schmidt}, K. and {Schmuckermaier}, F. and {Schweizer}, T. and {Sitarek}, J. and {{\v{S}}nidari{\'c}}, I. and {Sobczynska}, D. and {Spolon}, A. and {Stamerra}, A. and {Stri{\v{s}}kovi{\'c}}, J. and {Strom}, D. and {Strzys}, M. and {Suda}, Y. and {Suri{\'c}}, T. and {Takahashi}, M. and {Takeishi}, R. and {Tavecchio}, F. and {Temnikov}, P. and {Terzi{\'c}}, T. and {Teshima}, M. and {Tosti}, L. and {Truzzi}, S. and {Tutone}, A. and {Ubach}, S. and {van Scherpenberg}, J. and {Vanzo}, G. and {Vazquez Acosta}, M. and {Ventura}, S. and {Verguilov}, V. and {Viale}, I. and {Vigorito}, C.~F. and {Vitale}, V. and {Vovk}, I. and {Will}, M. and {Wunderlich}, C. and {Yamamoto}, T. and {Zari{\'c}}, D. and {Hodges}, M.},
        title = "{Investigating the Blazar TXS 0506+056 through Sharp Multiwavelength Eyes During 2017-2019}",
      journal = {\apj},
         year = 2022,
        month = mar,
       volume = {927},
       number = {2},
          eid = {197},
        pages = {197},
          doi = {10.3847/1538-4357/ac531d},
archivePrefix = {arXiv},
       eprint = {2202.02600},
 primaryClass = {astro-ph.HE},
       adsurl = {https://ui.adsabs.harvard.edu/abs/2022ApJ...927..197A}
}

@ARTICLE{2025A&A...700L..12K,
       author = {{Kovalev}, Y.~Y. and {Pushkarev}, A.~B. and {G{\'o}mez}, J.~L. and {Homan}, D.~C. and {Lister}, M.~L. and {Livingston}, J.~D. and {Pashchenko}, I.~N. and {Plavin}, A.~V. and {Savolainen}, T. and {Troitsky}, S.~V.},
        title = "{Looking into the jet cone of the neutrino-associated very high-energy blazar PKS 1424+240}",
      journal = {\aap},
         year = 2025,
        month = aug,
       volume = {700},
          eid = {L12},
        pages = {L12},
          doi = {10.1051/0004-6361/202555400},
archivePrefix = {arXiv},
       eprint = {2504.09287},
 primaryClass = {astro-ph.HE},
       adsurl = {https://ui.adsabs.harvard.edu/abs/2025A&A...700L..12K}
}

@ARTICLE{2025ApJ...991...33P,
       author = {{Plavin}, A.~V. and {Kovalev}, Y.~Y. and {Troitsky}, S.~V.},
        title = "{Extreme Jet Beaming Observed in Neutrino-associated Blazars}",
      journal = {\apj},
         year = 2025,
        month = sep,
       volume = {991},
       number = {1},
          eid = {33},
        pages = {33},
          doi = {10.3847/1538-4357/adf54f},
archivePrefix = {arXiv},
       eprint = {2503.08667},
 primaryClass = {astro-ph.HE},
       adsurl = {https://ui.adsabs.harvard.edu/abs/2025ApJ...991...33P}
}

@ARTICLE{2020ApJ...896L..19L,
       author = {{Lipunov}, V.~M. and {Kornilov}, V.~G. and {Zhirkov}, K. and {Gorbovskoy}, E. and {Budnev}, N.~M. and {Buckley}, D.~A.~H. and {Rebolo}, R. and {Serra-Ricart}, M. and {Podesta}, R. and {Tyurina}, N. and {Gress}, O. and {Sergienko}, Y. and {Yurkov}, V. and {Gabovich}, A. and {Balanutsa}, P. and {Gorbunov}, I. and {Vlasenko}, D. and {Balakin}, F. and {Topolev}, V. and {Pozdnyakov}, A. and {Kuznetsov}, A. and {Vladimirov}, V. and {Chasovnikov}, A. and {Kuvshinov}, D. and {Grinshpun}, V. and {Minkina}, E. and {Petkov}, V.~B. and {Svertilov}, S.~I. and {Lopez}, C. and {Podesta}, F. and {Levato}, H. and {Tlatov}, A. and {Van Soelen}, B. and {Razzaque}, S. and {B{\"o}ttcher}, M.},
        title = "{Optical Observations Reveal Strong Evidence for High-energy Neutrino Progenitor}",
      journal = {\apjl},
         year = 2020,
        month = jun,
       volume = {896},
       number = {2},
          eid = {L19},
        pages = {L19},
          doi = {10.3847/2041-8213/ab96ba},
archivePrefix = {arXiv},
       eprint = {2006.04918},
 primaryClass = {astro-ph.HE},
       adsurl = {https://ui.adsabs.harvard.edu/abs/2020ApJ...896L..19L}
}

@ARTICLE{2022MNRAS.510..469K,
       author = {{Kramarenko}, I.~G. and {Pushkarev}, A.~B. and {Kovalev}, Y.~Y. and {Lister}, M.~L. and {Hovatta}, T. and {Savolainen}, T.},
        title = "{A decade of joint MOJAVE-Fermi AGN monitoring: localization of the gamma-ray emission region}",
      journal = {\mnras},
         year = 2022,
        month = feb,
       volume = {510},
       number = {1},
        pages = {469-480},
          doi = {10.1093/mnras/stab3358},
archivePrefix = {arXiv},
       eprint = {2106.08416},
 primaryClass = {astro-ph.HE},
       adsurl = {https://ui.adsabs.harvard.edu/abs/2022MNRAS.510..469K}
}

@ARTICLE{2019MNRAS.486..430K,
       author = {{Kutkin}, A.~M. and {Pashchenko}, I.~N. and {Sokolovsky}, K.~V. and {Kovalev}, Y.~Y. and {Aller}, M.~F. and {Aller}, H.~D.},
        title = "{Opacity, variability, and kinematics of AGN jets}",
      journal = {\mnras},
         year = 2019,
        month = jun,
       volume = {486},
       number = {1},
        pages = {430-439},
          doi = {10.1093/mnras/stz885},
archivePrefix = {arXiv},
       eprint = {1809.05536},
 primaryClass = {astro-ph.GA},
       adsurl = {https://ui.adsabs.harvard.edu/abs/2019MNRAS.486..430K}
}

@ARTICLE{2019MNRAS.483L..42K,
       author = {{Kun}, E. and {Biermann}, P.~L. and {Gergely}, L. {\'A}.},
        title = "{Very long baseline interferometry radio structure and radio brightening of the high-energy neutrino emitting blazar TXS 0506+056}",
      journal = {\mnras},
         year = 2019,
        month = feb,
       volume = {483},
       number = {1},
        pages = {L42-L46},
          doi = {10.1093/mnrasl/sly216},
archivePrefix = {arXiv},
       eprint = {1807.07942},
 primaryClass = {astro-ph.HE},
       adsurl = {https://ui.adsabs.harvard.edu/abs/2019MNRAS.483L..42K}
}

@ARTICLE{2024MNRAS.527.8784A,
       author = {{Allakhverdyan}, V.~A. and {Avrorin}, A.~D. and {Avrorin}, A.~V. and {Aynutdinov}, V.~M. and {Barda{\v{c}}ov{\'a}}, Z. and {Belolaptikov}, I.~A. and {Bondarev}, E.~A. and {Borina}, I.~V. and {Budnev}, N.~M. and {Chadymov}, V.~A. and {Chepurnov}, A.~S. and {Dik}, V.~Y. and {Domogatsky}, G.~V. and {Doroshenko}, A.~A. and {Dvornick{\'y}}, R. and {Dyachok}, A.~N. and {Dzhilkibaev}, Zh-A.~M. and {Eckerov{\'a}}, E. and {Elzhov}, T.~V. and {Fajt}, L. and {Fomin}, V.~N. and {Gafarov}, A.~R. and {Golubkov}, K.~V. and {Gorshkov}, N.~S. and {Gress}, T.~I. and {Kebkal}, K.~G. and {Kharuk}, I. and {Khramov}, E.~V. and {Kolbin}, M.~M. and {Koligaev}, S.~O. and {Konischev}, K.~V. and {Korobchenko}, A.~V. and {Koshechkin}, A.~P. and {Kozhin}, V.~A. and {Kruglov}, M.~V. and {Kulepov}, V.~F. and {Lemeshev}, Y.~E. and {Milenin}, M.~B. and {Mirgazov}, R.~R. and {Naumov}, D.~V. and {Nikolaev}, A.~S. and {Petukhov}, D.~P. and {Pliskovsky}, E.~N. and {Rozanov}, M.~I. and {Ryabov}, E.~V. and {Safronov}, G.~B. and {Seitova}, D. and {Shaybonov}, B.~A. and {Shelepov}, M.~D. and {Shilkin}, S.~D. and {Shirokov}, E.~V. and {{\v{S}}imkovic}, F. and {Sirenko}, A.~E. and {Skurikhin}, A.~V. and {Solovjev}, A.~G. and {Sorokovikov}, M.~N. and {{\v{S}}tekl}, I. and {Stromakov}, A.~P. and {Suvorova}, O.~V. and {Tabolenko}, V.~A. and {Ulzutuev}, B.~B. and {Yablokova}, Y.~V. and {Zaborov}, D.~N. and {Zavyalov}, S.~I. and {Zvezdov}, D.~Y. and {Erkenov}, A.~K. and {Kosogorov}, N.~A. and {Kovalev}, Yu A. and {Kovalev}, Y.~Y. and {Plavin}, A.~V. and {Popkov}, A.~V. and {Pushkarev}, A.~B. and {Semikoz}, D.~V. and {Sotnikova}, Y.~V. and {Troitsky}, S.~V. and {(Baikal-GVD Collaboration)}},
        title = "{High-energy neutrino-induced cascade from the direction of the flaring radio blazar TXS 0506+056 observed by Baikal-GVD in 2021}",
      journal = {\mnras},
         year = 2024,
        month = jan,
       volume = {527},
       number = {3},
        pages = {8784-8792},
          doi = {10.1093/mnras/stad3653},
archivePrefix = {arXiv},
       eprint = {2210.01650},
 primaryClass = {astro-ph.HE},
       adsurl = {https://ui.adsabs.harvard.edu/abs/2024MNRAS.527.8784A}
}

@ARTICLE{2021ApJ...923...67H,
       author = {{Homan}, D.~C. and {Cohen}, M.~H. and {Hovatta}, T. and {Kellermann}, K.~I. and {Kovalev}, Y.~Y. and {Lister}, M.~L. and {Popkov}, A.~V. and {Pushkarev}, A.~B. and {Ros}, E. and {Savolainen}, T.},
        title = "{MOJAVE. XIX. Brightness Temperatures and Intrinsic Properties of Blazar Jets}",
      journal = {\apj},
         year = 2021,
        month = dec,
       volume = {923},
       number = {1},
          eid = {67},
        pages = {67},
          doi = {10.3847/1538-4357/ac27af},
archivePrefix = {arXiv},
       eprint = {2109.04977},
 primaryClass = {astro-ph.HE},
       adsurl = {https://ui.adsabs.harvard.edu/abs/2021ApJ...923...67H}
}

@ARTICLE{BK79,
       author = {{Blandford}, R.~D. and {K{\"o}nigl}, A.},
        title = "{Relativistic jets as compact radio sources.}",
      journal = {\apj},
         year = 1979,
        month = aug,
       volume = {232},
        pages = {34-48},
          doi = {10.1086/157262},
       adsurl = {https://ui.adsabs.harvard.edu/abs/1979ApJ...232...34B}
}

@ARTICLE{2018ApJ...854L..32P,
       author = {{Paiano}, Simona and {Falomo}, Renato and {Treves}, Aldo and {Scarpa}, Riccardo},
        title = "{The Redshift of the BL Lac Object TXS 0506+056}",
      journal = {\apjl},
         year = 2018,
        month = feb,
       volume = {854},
       number = {2},
          eid = {L32},
        pages = {L32},
          doi = {10.3847/2041-8213/aaad5e},
archivePrefix = {arXiv},
       eprint = {1802.01939},
 primaryClass = {astro-ph.GA},
       adsurl = {https://ui.adsabs.harvard.edu/abs/2018ApJ...854L..32P}
}

@ARTICLE{2018ApJS..234...12L,
       author = {{Lister}, M.~L. and {Aller}, M.~F. and {Aller}, H.~D. and {Hodge}, M.~A. and {Homan}, D.~C. and {Kovalev}, Y.~Y. and {Pushkarev}, A.~B. and {Savolainen}, T.},
        title = "{MOJAVE. XV. VLBA 15 GHz Total Intensity and Polarization Maps of 437 Parsec-scale AGN Jets from 1996 to 2017}",
      journal = {\apjs},
         year = 2018,
        month = jan,
       volume = {234},
       number = {1},
          eid = {12},
        pages = {12},
          doi = {10.3847/1538-4365/aa9c44},
archivePrefix = {arXiv},
       eprint = {1711.07802},
 primaryClass = {astro-ph.GA},
       adsurl = {https://ui.adsabs.harvard.edu/abs/2018ApJS..234...12L}
}

@ARTICLE{MOJAVE_XIV,
       author = {{Pushkarev}, A.~B. and {Kovalev}, Y.~Y. and {Lister}, M.~L. and {Savolainen}, T.},
        title = "{MOJAVE - XIV. Shapes and opening angles of AGN jets}",
      journal = {\mnras},
         year = 2017,
        month = jul,
       volume = {468},
       number = {4},
        pages = {4992-5003},
          doi = {10.1093/mnras/stx854},
archivePrefix = {arXiv},
       eprint = {1705.02888},
 primaryClass = {astro-ph.HE},
       adsurl = {https://ui.adsabs.harvard.edu/abs/2017MNRAS.468.4992P}
}

@ARTICLE{2009ApJS..180..330K,
       author = {{Komatsu}, E. and {Dunkley}, J. and {Nolta}, M.~R. and {Bennett}, C.~L. and {Gold}, B. and {Hinshaw}, G. and {Jarosik}, N. and {Larson}, D. and {Limon}, M. and {Page}, L. and {Spergel}, D.~N. and {Halpern}, M. and {Hill}, R.~S. and {Kogut}, A. and {Meyer}, S.~S. and {Tucker}, G.~S. and {Weiland}, J.~L. and {Wollack}, E. and {Wright}, E.~L.},
        title = "{Five-Year Wilkinson Microwave Anisotropy Probe Observations: Cosmological Interpretation}",
      journal = {\apjs},
         year = 2009,
        month = feb,
       volume = {180},
       number = {2},
        pages = {330-376},
          doi = {10.1088/0067-0049/180/2/330},
archivePrefix = {arXiv},
       eprint = {0803.0547},
 primaryClass = {astro-ph},
       adsurl = {https://ui.adsabs.harvard.edu/abs/2009ApJS..180..330K}
}

@ARTICLE{OVRO_IC_21,
       author = {{Hovatta}, T. and {Lindfors}, E. and {Kiehlmann}, S. and {Max-Moerbeck}, W. and {Hodges}, M. and {Liodakis}, I. and {L{\"a}hteem{\"a}ki}, A. and {Pearson}, T.~J. and {Readhead}, A.~C.~S. and {Reeves}, R.~A. and {Suutarinen}, S. and {Tammi}, J. and {Tornikoski}, M.},
        title = "{Association of IceCube neutrinos with radio sources observed at Owens Valley and Mets{\"a}hovi Radio Observatories}",
      journal = {\aap},
         year = 2021,
        month = jun,
       volume = {650},
          eid = {A83},
        pages = {A83},
          doi = {10.1051/0004-6361/202039481},
archivePrefix = {arXiv},
       eprint = {2009.10523},
 primaryClass = {astro-ph.HE},
       adsurl = {https://ui.adsabs.harvard.edu/abs/2021A&A...650A..83H}
}

@ARTICLE{Daly_Marscher_1988,
       author = {{Daly}, Ruth A. and {Marscher}, Alan P.},
        title = "{The Gasdynamics of Compact Relativistic Jets}",
      journal = {Astrophys. J.},
         year = 1988,
        month = nov,
       volume = {334},
        pages = {539},
          doi = {10.1086/166858},
       adsurl = {https://ui.adsabs.harvard.edu/abs/1988ApJ...334..539D}
}

@ARTICLE{Polina,
       author = {{Kalashev}, Oleg and {Kivokurtseva}, Polina and {Troitsky}, Sergey},
        title = "{Neutrino production in blazar radio cores}",
      journal = {\jcap},
         year = 2023,
        month = dec,
       volume = {2023},
       number = {12},
          eid = {007},
        pages = {007},
          doi = {10.1088/1475-7516/2023/12/007},
archivePrefix = {arXiv},
       eprint = {2212.03151},
 primaryClass = {astro-ph.HE},
       adsurl = {https://ui.adsabs.harvard.edu/abs/2023JCAP...12..007K}
}

@ARTICLE{2020AdSpR..65..745K,
       author = {{Kovalev}, Yu. A. and {Kardashev}, N.~S. and {Kovalev}, Y.~Y. and {Sokolovsky}, K.~V. and {Voitsik}, P.~A. and {Edwards}, P.~G. and {Popkov}, A.~V. and {Zhekanis}, G.~V. and {Sotnikova}, Yu. V. and {Nizhelsky}, N.~A. and {Tsybulev}, P.~G. and {Erkenov}, A.~K. and {Bursov}, N.~N.},
        title = "{RATAN-600 and RadioAstron reveal the neutrino-associated blazar TXS 0506+056 as a typical variable AGN}",
      journal = {Advances in Space Research},
         year = 2020,
        month = jan,
       volume = {65},
       number = {2},
        pages = {745-755},
          doi = {10.1016/j.asr.2019.04.034},
       adsurl = {https://ui.adsabs.harvard.edu/abs/2020AdSpR..65..745K}
}

@ARTICLE{1999A&AS..139..545K,
       author = {{Kovalev}, Y.~Y. and {Nizhelsky}, N.~A. and {Kovalev}, Yu. A. and {Berlin}, A.~B. and {Zhekanis}, G.~V. and {Mingaliev}, M.~G. and {Bogdantsov}, A.~V.},
        title = "{Survey of instantaneous 1-22 GHz spectra of 550 compact extragalactic objects with declinations from -30$^{deg}$ to +43$^{deg}$}",
      journal = {\aaps},
         year = 1999,
        month = nov,
       volume = {139},
        pages = {545-554},
          doi = {10.1051/aas:1999406},
archivePrefix = {arXiv},
       eprint = {astro-ph/0408264},
 primaryClass = {astro-ph},
       adsurl = {https://ui.adsabs.harvard.edu/abs/1999A&AS..139..545K}
}

@ARTICLE{neutrino-from-spine-sheath,
       author = {{Tavecchio}, F. and {Ghisellini}, G.},
        title = "{High-energy cosmic neutrinos from spine-sheath BL Lac jets}",
      journal = {\mnras},
         year = 2015,
        month = aug,
       volume = {451},
       number = {2},
        pages = {1502-1510},
          doi = {10.1093/mnras/stv1023},
archivePrefix = {arXiv},
       eprint = {1411.2783},
 primaryClass = {astro-ph.HE},
       adsurl = {https://ui.adsabs.harvard.edu/abs/2015MNRAS.451.1502T}
}

@ARTICLE{0506sheath,
       author = {{Ros}, E. and {Kadler}, M. and {Perucho}, M. and {Boccardi}, B. and {Cao}, H. -M. and {Giroletti}, M. and {Krau{\ss}}, F. and {Ojha}, R.},
        title = "{Apparent superluminal core expansion and limb brightening in the candidate neutrino blazar TXS 0506+056}",
      journal = {\aap},
         year = 2020,
        month = jan,
       volume = {633},
          eid = {L1},
        pages = {L1},
          doi = {10.1051/0004-6361/201937206},
archivePrefix = {arXiv},
       eprint = {1912.01743},
 primaryClass = {astro-ph.GA},
       adsurl = {https://ui.adsabs.harvard.edu/abs/2020A&A...633L...1R}
}

@ARTICLE{2021ApJ...923...30L,
       author = {{Lister}, M.~L. and {Homan}, D.~C. and {Kellermann}, K.~I. and {Kovalev}, Y.~Y. and {Pushkarev}, A.~B. and {Ros}, E. and {Savolainen}, T.},
        title = "{Monitoring Of Jets in Active Galactic Nuclei with VLBA Experiments. XVIII. Kinematics and Inner Jet Evolution of Bright Radio-loud Active Galaxies}",
      journal = {\apj},
         year = 2021,
        month = dec,
       volume = {923},
       number = {1},
          eid = {30},
        pages = {30},
          doi = {10.3847/1538-4357/ac230f},
archivePrefix = {arXiv},
       eprint = {2108.13358},
 primaryClass = {astro-ph.HE},
       adsurl = {https://ui.adsabs.harvard.edu/abs/2021ApJ...923...30L}
}

@ARTICLE{2020ApJ...891..115P,
       author = {{Petropoulou}, Maria and {Murase}, Kohta and {Santander}, Marcos and {Buson}, Sara and {Tohuvavohu}, Aaron and {Kawamuro}, Taiki and {Vasilopoulos}, Georgios and {Negoro}, Hiroshi and {Ueda}, Yoshihiro and {Siegel}, Michael H. and {Keivani}, Azadeh and {Kawai}, Nobuyuki and {Mastichiadis}, Apostolos and {Dimitrakoudis}, Stavros},
        title = "{Multi-epoch Modeling of TXS 0506+056 and Implications for Long-term High-energy Neutrino Emission}",
      journal = {\apj},
         year = 2020,
        month = mar,
       volume = {891},
       number = {2},
          eid = {115},
        pages = {115},
          doi = {10.3847/1538-4357/ab76d0},
archivePrefix = {arXiv},
       eprint = {1911.04010},
 primaryClass = {astro-ph.HE},
       adsurl = {https://ui.adsabs.harvard.edu/abs/2020ApJ...891..115P}
}

@ARTICLE{2025ApJS..276...38P,
       author = {{Petrov}, L.~Y. and {Kovalev}, Y.~Y.},
        title = "{The Radio Fundamental Catalog. I. Astrometry}",
      journal = {\apjs},
         year = 2025,
        month = feb,
       volume = {276},
       number = {2},
          eid = {38},
        pages = {38},
          doi = {10.3847/1538-4365/ad8c36},
archivePrefix = {arXiv},
       eprint = {2410.11794},
 primaryClass = {astro-ph.IM},
       adsurl = {https://ui.adsabs.harvard.edu/abs/2025ApJS..276...38P}
}
\end{document}